\documentclass[structabstract]{aa}  

\usepackage{graphicx}
\usepackage{txfonts}
\usepackage[usenames]{color}

\usepackage{natbib,twoopt}
\bibpunct{(}{)}{;}{a}{}{,} 
\newcommandtwoopt{\citeads}[3][][]{\href{http://adsabs.harvard.edu/abs/#3}%
{\citealp[#1][#2]{#3}}}
\newcommandtwoopt{\citepads}[3][][]{\href{http://adsabs.harvard.edu/abs/#3}%
{\citep[#1][#2]{#3}}}
\newcommandtwoopt{\citetads}[3][][]{\href{http://adsabs.harvard.edu/abs/#3}%
{\citet[#1][#2]{#3}}} 
\newcommandtwoopt{\citeyearads}[3][][]%
{\href{http://adsabs.harvard.edu/abs/#3}{\citeyear[#1][#2]{#3}}} 

\begin{document}

   \title{The physical state of selected cold clumps\thanks{{\it Herschel} is an ESA space observatory with science instruments provided by European-led Principal Investigator consortia and with important participation from NASA.}}

   \subtitle{}

   \author{A. Parikka \inst{1,2}
          \and
          M. Juvela \inst{1}
          \and
          V.-M. Pelkonen \inst{1,3}
          \and
          J. Malinen \inst{1}
          \and
          J. Harju \inst{1}
          }

   \institute{Department of Physics, P.O.Box 64, FI-00014, University of Helsinki, Finland
   \and Institut d'Astrophysique Spatiale, Universit\'e Paris Sud, 91405 Orsay Cedex, France 
   \and Finnish Centre for Astronomy with ESO (FINCA), University of Turku, V\"ais\"al\"antie 20, 21500 Piikki\"o, Finland }

   \date{Received <date> / Accepted <date>}

 
  \abstract
   {The study of prestellar cores is essential to understanding the initial stages of star formation. With $Herschel$ more cold clumps have been detected than ever before. For this study we have selected 21 cold clumps from 20 $Herschel$ fields observed as a follow-up on original $Planck$ detections. We have observed these clumps in $^{13}$CO (1-0), C$^{18}$O (1-0), and N$_2$H$^+$ (1-0) lines.}
   {Our aim is to find out if these cold clumps are prestellar. We have examined to what extent independent analysis of the dust and the molecular lines lead to similar conclusions about the masses of these objects.}
   {We calculate the clump masses and densities from the dust continuum and molecular line observations and compare these to each other and to the virial and Bonnor-Ebert masses calculated for each clump. Finally we examine two of the fields with radiative transfer models to estimate CO abundances.}
   {When excitation temperatures could be estimated, the column densities derived from molecular line observations were comparable to those from dust continuum data. The median column density estimates are 4.2$\times 10^{21}$cm$^{-2}$ and 5.5$\times 10^{21}$cm$^{-2}$ for the line and dust emission data, respectively. The calculated abundances, column densities, volume densities, and masses all have large uncertainties and one must be careful when drawing conclusions. Abundance of $^{13}$CO was found in modeling  the two clumps in the field G131.65$+$9.75 to be close to the usual value of 10$^{-6}$. The abundance ratio of $^{13}$CO and C$^{18}$O was $\sim$10. Molecular abundances could only be estimated  with modeling, relying on dust column density data.}
   {The results indicate that most cold clumps, even those with dust color temperatures close to 11 K, are not necessarily prestellar.}

   \keywords{ISM: clouds --
                submillimeter: ISM, dust, molecular lines --
                stars: formation
               }

   \maketitle
%

\section{Introduction}

The details of the initial stages of star formation are still poorly known, despite many observational and theoretical studies. This is in part because star formation involves a very complex mixture of effects of gravity, turbulence, rotation, radiation, thermodynamics, and magnetic fields. Both large representative samples of clumps and detailed studies of their properties and relations to their cloud environment are needed.

$Herschel$ observations of star-forming clouds, e.g., those conducted within the Gould Belt Survey \citep{Andre2010} and HOBYS program \citep{Motte2010}, have lead to the detections of hundreds of starless and protostellar condensations. This is particularly true for nearby star forming clouds where the $Herschel$ resolution is sufficient to resolve gravitationally bound cores. The objects appear preferentially within filaments, which are a conspicuous feature of all interstellar clouds \citep{Andre2013}. The filaments themselves contain both supercritical gravitationally bound structures and subcritical filaments that are likely to disperse with time \citep{Arzoumanian2013}. 

The $Planck$ satellite has performed an all-sky survey to map anisotropies of the cosmic microwave background \citep{Tauber2010a}. At the same time $Planck$ is providing maps of thermal dust emission from molecular clouds within the Milky Way. From these data more than 10,000 compact cold sources have been detected and a large fraction of these are believed to be prestellar cores or larger structures harboring prestellar cores \citep{PlanckCollaboration2011c}. The $Planck$ and IRAS (100~$\muup$m) surveys have been used to compose the Cold Clump Catalogue of $Planck$ Objects (C3PO),  from which 915 most reliable detections were published as the Early Cold Clumps Catalogue (353$-$857~GHz, 350$-$850~$\muup$m), ECC \citep[see, e.g.,][] {PlanckCollaboration2011b, PlanckCollaboration2011e, PlanckCollaboration2011d}. The sources were detected with the method described in \citet{Montier2010} and the initial results on the Planck detections were described in \citet{PlanckCollaboration2011d, PlanckCollaboration2011c}. The final version of C3PO is currently in preparation and will provide a global view of the cold clump population over the whole sky.

A number of $Planck$ detections of cold clumps have been confirmed with the $Herschel$ Space Observatory observations (100$-$500~$\muup$m) to be cold (T$_{\rm dust} \sim$14~K or below) and also dense. For the present study, we selected clumps  from fields covered in the $Herschel$ open time key program Galactic Cold Cores \citep{Juvela2010}. The aim of this $Herschel$ program is to examine a representative cross section of the source population of cold clumps observed with $Planck$ and to determine the physical properties of these clumps. The $Herschel$ results suggest that many of the sources are already past the prestellar phase \citep{Juvela2010, Juvela2011}. In this paper, we refer to sources that are gravitationally bound as prestellar objects. If the estimated mass exceeds the virial mass, the object is expected to be gravitationally bound. Bonnor-Ebert (BE) spheres are used as an alternative model to recognize prestellar cores \citep{Andre2010, Konyves2010a}.

In an earlier study, a sample of 71 fields at distances ranging from $\sim$100 pc to several kiloparsecs were examined. The $Herschel$ observations, together with AKARI and WISE infrared data, were used to confirm the presence of cold dust. In about half of the observed fields, point sources were found in the mid-infrared, indicating active star formation. However, the cold dust still dominated the submillimeter spectra in these active star formation areas \citep{Juvela2012a}. 

Some of the $Planck$-detected clumps have already been  studied in molecular
lines. \citet{Wu2012} surveyed 673 sources with single-point observations and mapped 10 clumps with 22 identified potential cold cores. They found seven cores that are likely to be gravitationally bound and, thus, on the verge of collapse. In a follow-up study, 71 of the sources observed only with single-point observations were examined and 90 \% of the found cores were starless \citep{Meng2013}. The clumps studied by \citet{Wu2012} and \citet{Meng2013} did not include the clumps chosen for this paper.

The definition of dense structures in the ISM is still an ongoing process. The clumps are small and dense condensations in the molecular clouds, often inside filaments. Cores are even smaller and denser. \citet{Bergin2007} defined the clumps to have a mass of 50$-$500~M$_\odot$, size of 0.3$-$3~pc, and density of 10$^3-$10$^4$~cm$^{-3}$. Cores were defined to have mass of 0.5$-$5~M$_\odot$, size of 0.03$-$0.2~pc, and density of 10$^4-$10$^5$~cm$^{-3}$. We use the word clump as a general term for the dense structures we investigate here, however, some  of these structures are likely to be cores themselves or contain unresolved cores.

Extensive surveys have been done in $^{12}$CO and $^{13}$CO over the Galactic plane and of nearby star-forming clouds (see, e.g., \citealp{Combes1991} for review). However,  general surveys are not the best method to research possible compact sources. For the first time, with $Planck$ and $Herschel$, we can identify the coldest and densest clumps in the known molecular clouds complexes and focus on dedicated molecular line observations targeting the most interesting dense clumps in these regions.


We combine $Herschel$ data with molecular line observations to examine the physical properties of several cold clumps and their likelihood to evolve into star-forming cores. To investigate the gravitational stability of the objects, molecular line observations are essential to providing direct information of the kinetic and turbulent forces and, thus, on the current stability of the clumps and the cores. The objects can become gravitationally unstable by cooling and accretion, possibly aided or hindered by turbulence and external forces \citep{Bergin2007, Ballesteros-Paredes2007}.

We investigate the physical state of a few clumps that were originally selected based on the cold dust signature detected by the $Planck$ satellite. We derive the CO column densities, volume densities, and clump masses  and estimate the gravitational stability. The potential prestellar nature of the objects is investigated by comparing their mass estimates with virial masses. We look for signs of CO abundance variations as further evidence of the nature of the sources. Analysis assuming a local thermodynamic equilibrium (LTE) will be complemented with modeling. We selected one target for radiative transfer modeling where continuum data are used to constrain the models so that direct estimates of CO abundance are possible. The results also provide insights into the nature of the larger $Planck$ cold clump population.

The paper is organized as follows. The observations are described in Sect. \ref{observations}. Further, we go through the methods in Sect. \ref{methods}. The results are presented in Sect. \ref{results} and they are discussed in more detail in Sect. \ref{discussion}. Finally, the conclusions are summarized in Sect. \ref{conclusions}.


\section{Observations}
\label{observations}


\subsection{The clump sample}

We examine 21 clumps in 20 fields previously observed with $Herschel$ at wavelengths 100~-~500~$\muup$m. The fields mapped with $Herschel$ were originally selected based on the $Planck$ survey, in which these clumps showed a significant excess of cold dust emission \citep{PlanckCollaboration2011b}. Our selection was based on the $Herschel$ data from the  {\em Galactic Cold Cores} $Herschel$ key program (PI Juvela) that carried out follow-up observations of 116 fields, each with one or more $Planck$ cold clumps. We chose the bright clumps, which based on $Herschel$ data also contain very cold dust (color temperature at $T_{\rm dust} \la$14~K, 40$\arcsec$ resolution). The fields are located in the Milky Way at galactic latitudes $| b |$=$9-17\degr$ , which ensures minimal confusion from background emission. While they typically have a general cometary or filamentary morphology, the submillimeter data show that at small scales the clouds are fragmented to several clumps.

\subsection{Molecular line observations}

We mapped the selected clumps  in the $^{13}$CO (1-0) spectral line using the 20-m radio telescope in Onsala Space Observatory, Sweden, in February and March 2012. Toward the $^{13}$CO peaks, we also observed C$^{18}$O (1-0) and N$_2$H$^+$ (1-0) lines. Part of the northern clump in field G131.65+9.75 was mapped in $^{13}$CO (1-0) in February 2011 \citep{PlanckCollaboration2011c}. The observed clumps are summarized in Table \ref{sources} and the observed spectra are shown in the appendix \ref{co_spectra_app}.
\begin{table*}
        \caption{Summary of observations. The table lists the 21 targets, kinematic distances for the fields \citep{Montillaud}, temperature of the dust, coordinates for the $^{13}$CO (1-0) peak temperature (where the C$^{18}$O (1-0) and N$_2$H$^+$ (1-0) were observed), area in square arc minutes selected for the $^{13}$CO (1-0) mapping, and number of observed points for C$^{18}$O (1-0) and N$_2$H$^+$ (1-0).}
        \vspace{0,2cm}
        \centering
        \begin{tabular}{ |l|c|c|c|c|c|c|c| }
                \hline
        Field & $d$ (pc) & $T_{\rm dust}$ & $\alpha$ (J2000) & $\delta$ (J2000) & $^{13}$CO (sq$\arcmin$) & C$^{18}$O (N) & N$_2$H$^+$ (N) \\
                \hline
                G86.97-4.06    & 700  & 11.5$\pm$0.9 & 21 17 44.22 & +43 18 47.1 & 4.55    & 1 & 1 \\
                G92.04+3.93    & 800  & 10.4$\pm$0.8 & 21 02 23.43 & +52 28 38.3 & 3.96    & 1 & 1 \\
                G93.21+9.55    & 440  & 10.6$\pm$0.8 & 20 37 19.98 & +56 54 42.6 & 6.79    & 2 & 1 \\
                G94.15+6.50    & 800  & 12$\pm$1     & 20 59 04.80 & +55 34 33.4 & 6.67    & 1 & 1 \\
                G98.00+8.75    & 1100 & 12$\pm$1     & 21 04 22.70 & +60 08 55.1 & 3.79    & 1 & 1 \\
                G105.57+10.39  & 900  & 10.9$\pm$0.8 & 21 41 01.79 & +66 34 40.9 & 8.33    & 1 & 1 \\
                G108.28+16.68  & 300  & 14$\pm$1     & 21 09 08.60 & +72 53 43.2 & 8.01    & 7 & 1 \\
                G110.80+14.16  & 400  & 14$\pm$1     & 21 57 35.84 & +72 46 49.1 & 6.88    & 2 & 1 \\
                G111.41-2.95   & 3000 & 13$\pm$2     & 23 22 28.78 & +57 36 44.5 & 7.77    & 1 & 1 \\
                G131.65+9.75 N & 1070 & 11$\pm$1     & 02 40 03.30 & +70 43 20.1 & 6.79    & 4 & 1 \\
                G131.65+9.75 S & 1070 & 13$\pm$1     & 02 40 11.20 & +70 36 09.5 & 8.89    & 7 & 1 \\
                G132.12+8.95   & 850  & 10.6$\pm$0.9 & 02 41 49.64 & +69 50 12.9 & 5.92    & 2 & 1 \\
                G149.67+3.56   & 170  & 13$\pm$1     & 04 17 08.20 & +55 13 40.6 & 4.89    & 1 & 1 \\
                G154.08+5.23   & 170  & 11.2$\pm$0.9 & 04 48 08.73 & +53 07 23.1 & 20.2    & 3 & 1 \\
                G157.92-2.28   & 2500 & 10.9$\pm$0.8 & 04 28 48.30 & +45 24 22.5 & 8.33    & 2 & 1 \\
                G159.34+11.21  & 1590 & 14$\pm$1     & 05 42 17.30 & +52 08 16.4 & 1 point & 1 & 1 \\
                G161.55-9.30   & 250  & 13$\pm$1     & 04 16 15.30 & +37 45 35.0 & 2.22    & 5 & 1 \\
                G164.71-5.64   & 330  & 13$\pm$1     & 04 40 44.43 & +37 45 14.3 & 5.39    & 1 & 1 \\
                G167.20-8.69   & 160  & 12$\pm$1     & 04 35 26.88 & +34 18 25.6 & 6.88    & 2 & 1 \\
                G168.85-10.19  & 2610 & 13$\pm$1     & 04 37 07.59 & +31 43 01.2 & 1 point & 1 & 0 \\
                G173.43-5.44   & 150  & 13$\pm$1     & 05 07 49.46 & +31 20 46.3 & 1 point & 1 & 0 \\
                \hline 
        \end{tabular}
        \label{sources}
\end{table*}
We carried out the observations  in frequency switching mode, with frequency throws of $\pm$10~MHz for $^{13}$CO and C$^{18}$O and $\pm$5~MHz for N$_2$H$^+$. For N$_2$H$^+$ in the 86 GHz band the half-power beam width, HPBW, is 44$\arcsec$ and the main beam effciency, $\eta_{\rm B}$, is 0.65. For $^{13}$CO and C$^{18}$O in the 109 GHz band the HPBW is 35$\arcsec$ and $\eta_{\rm B}$ is 0.45.

The telescope has a pointing accuracy of 3$\arcsec$ rms. The pointing was checked with SiO spectra toward bright sources R Cassiopeiae and U Orionis. The calibration was achieved through the chopper-wheel method, and the system gives antenna temperatures in units of $T_{\rm A}^*$ \citep{Kutner1981}. The backend was a low-resolution digital autocorrelator spectrometer (ACS) with a bandwidth of 40 MHz divided into 1198 channels. This gave a channel width of 25 kHz, corresponding to 0.068 km~s$^{-1}$ at the frequency of the $^{13}{\rm CO}(1-0)$ line.

On average the RMS noise for the observed $^{13}$CO lines was $\sim$0.27 K per channel on the main beam temperature ($T_{\rm MB}$) scale. The corresponding values for the observed ${\rm C^{18}O}$ and ${\rm N_2H^+}$ lines are $\sim$0.06 K and $\sim$0.05 K, respectively. The data were reduced with GILDAS/CLASS software package\footnote{Institut de Radioastronomie Millim\'etrique (IRAM):\\ http://iram.fr/IRAMFR/GILDAS/}. The frequency switched spectra were folded, averaged, and the spectral baselines were modeled and subtracted using third order polynomials, except for C$^{18}$O in sources G108.28+16.68, G131.65+9.75 north and south, and G161.55-9.30 where fifth order polynomials were used.


\subsection{Dust continuum observations with $Herschel$}

We observed the dust continuum data with the $Herschel$ SPIRE instrument \citep{Griffin2010} between November 2009 and May 2011. These observations consist of  surface brightness maps of 250, 350, and 500 $\muup$m. The raw and pipeline-reduced data are available via the {\it Herschel} Science Archive, the user-reduced maps are available via ESA site\footnote{\em http://herschel.esac.esa.int/UserReducedData.shtml}

We reduced these data  similar to \citep{Juvela2012a} and only a summary of these reduction steps is given here. We completed the reduction  using the $Herschel$ Interactive Processing Environment HIPE v.10.0 and the official pipeline with the iterative destriper and extended emission calibration options. The resulting maps are the product of direct projection onto the sky and averaging of the time-ordered data. The resolutions of the maps are 18$\arcsec$, 25$\arcsec$, and 37$\arcsec$ at 250, 350, and 500~$\muup$m, respectively. 

The accuracy of the gain calibration of $Herschel$ data is better than 7 \% in absolute terms and probably better than 2~\% band-to-band \footnote{SPIRE Observer's manual\\ \em http://herschel.esac.esa.int/Documentation.shtml}. The surface brightness data do not have an absolute zero point and therefore, before temperatures or column densities can be estimated, we must set a consistent zero point across all $Herschel$ bands. One alternative is to compare $Herschel$ data with $Planck$ and IRAS measurements (using IRAS data tied to DIRBE scale) and to make use of the zero points of those surveys \citep[see, e.g.,][]{Juvela2012a}. We opt for the more straightforward procedure of selecting a reference area with a radius of 1.5$\arcmin$ within the $Herschel$ map and measuring surface brightness values relative to the average value found in the reference area. The central coordinates of the reference areas can be seen in Table \ref{ref_areas}.
\begin{table}
        \caption{Center coordinates of the 1.5$\arcmin$ radius reference areas.}
        \vspace{0,2cm}
        \centering
        \begin{tabular}{ |l|c|c| }
                \hline
                Field & $\alpha$ (J2000) & $\delta$ (J2000) \\
                \hline
        G86.97-4.06     &  21 16 40.8 & +43 34 45   \\
        G92.04+3.93     &  21 01 17.9 & +52 34 45   \\
        G93.21+9.55     &  20 37 44.3 & +56 44 49   \\
        G94.15+6.50     &  20 59 48.7 & +55 56 19   \\
        G98.00+8.75     &  21 05 59.4 & +59 51 19   \\
        G105.57+10.39   &  21 43 14.5 & +66 26 12   \\
        G108.28+16.68   &  21 08 16.0 & +72 51 20   \\
        G110.80+14.16   &  21 59 05.7 & +72 37 40   \\
        G111.41-2.95    &  23 20 11.9 & +57 35 20   \\
        G131.65+9.75    &  02 42 10.5 & +70 45 20   \\
        G132.12+8.95    &  02 37 04.7 & +69 53 43   \\
        G149.67+3.56    &  04 18 42.2 & +55 29 55   \\
        G154.08+5.23    &  04 47 58.8 & +53 14 42   \\
        G157.92-2.28    &  04 30 02.5 & +45 25 50   \\
        G159.34+11.21   &  05 41 07.7 & +51 54 33   \\
        G161.55-9.30    &  04 15 46.2 & +37 40 56   \\
        G164.71-5.64    &  04 42 47.6 & +38 22 45   \\
        G167.20-8.69    &  04 35 10.1 & +34 11 09   \\
        G168.85-10.19   &  04 36 06.8 & +31 50 54   \\
        G173.43-5.44    &  05 07 55.2 & +31 04 55   \\
                \hline 
        \end{tabular}
        \label{ref_areas}
\end{table}
The derived dust temperature and column density estimates ignore the very diffuse medium to the extent that is visible in the reference region. The reference regions have low extinction ($A_{\rm V}\sim 2$ or less) and therefore probably a lower fractional abundance of CO. Thus, even after background subtraction, the continuum data may probe a volume larger than the actual molecular cloud. Therefore, the extrapolation of the relations like in Fig. \ref{comp} does not necessarily go via the origin of that plot. This should not be a significant problem, however, because we are interested in the densest clumps that are far above the column density of the reference regions.

The resolution of $Herschel$ column density maps is 40$\arcsec$. The beam size is 35$\arcsec$ in CO observations and 40$\arcsec$ in ${\rm N_2H^+}$ observations. Because of the relatively small differences and the extended nature of our sources, the data were compared directly without further convolution.


\section{Methods}
\label{methods}

For the calculation of the CO column densities, we used the method described by \citet{Myers1983}. Based on the data analysis, we calculated the optical depth at the peak of C$^{18}$O line, $\tau_{18}$, the excitation temperature of the C$^{18}$O line, $T_{18}$, and the column density of the C$^{18}$O in the cloud, $N_{18}$. To derive the total column density of H$_2$, the column density of C$^{18}$O, $N_{18}$, is divided by the commonly assumed abundance ratio [C$^{18}$O]/[H$_2$]=$10^{-7}$.

The optical depths of $^{13}$CO and C$^{18}$O are related by (\citet{Myers1983}, Eq. 2)
\begin{equation} 
        \tau_{13} \, = \, \tau_{18} \, \frac{n_{13} (J=1)}{n_{18} (J=1)} \, \frac{L_{13}}{L_{18}} \, \frac{\Delta V_{18}}{\Delta V_{13}} \, \frac{J(T_{18})}{J(T_{13})}.
\end{equation}
In the equation $J(T) \, = \, T_0 [\exp (T_0/T)-1]^{-1}$, where $T_0$ = 5.27~K for C$^{18}$O and 5.29~K for $^{13}$CO and $T$ is the excitation temperature. The $n_{13}$(J=1) and $n_{18}$(J=1) are the number of molecules at J=1 levels of the $^{13}$CO J=1-0 and C$^{18}$O, respectively. The $L_{13}$ and $L_{18}$ are the line-of-sight extent of the emitting gas and the $\Delta V_{13}$ and $\Delta V_{18}$ are the line widths of the $^{13}$CO and C$^{18}$O line, respectively. The equation assumes the same excitation temperature for both lines, the same beam filling (both lines originate in the same region), and the same velocity gradient. We also assume the terrestrial abundance ratio of 5.5 between $^{13}$CO and C$^{18}$O and we use the line width of C$^{18}$O for the calculations. For a more comprehensive explanation of the method, see \citet{Myers1983}.


We also derived the column density with the $Herschel$ dust continuum observations. The observed intensity can be stated as\begin{equation}
        I_\nu \, = \, B_\nu(T_{\rm dust})(1-e^{-\tau}) \, \approx \, B(T_{\rm dust}) \times \tau
        \label{eq:int}
,\end{equation}
where $I_\nu$ is the observed intensity at a frequency $\nu$, $B_\nu(T_{\rm dust})$ is the blackbody brightness of the object as a function of color temperature $T_{\rm dust}$ , and $\tau$ is the source optical depth. The equation assumes a homogeneous source. In the far-infrared and at longer wavelengths the optical depths of the clouds are clearly below one. This justifies the approximation made in Eq. \ref{eq:int}. The optical depth can be written as
\begin{equation}
        \tau \, = \, \kappa_\nu \times N({\rm H_2}) \times \mu
        \label{eq:op_depth}
,\end{equation}
where $\muup$ is the average particle mass per H$_2$ molecule, 2.8 u. We assume a dust opacity of $\kappa = 0.1(f/1000.0 \times 10^9)^{2.0} \, \rm{cm}^2/g$, where $f$ is the frequency \citep{Beckwith1990}. The constant 2.0 is the dust opacity spectral index, $\beta$. While it may vary from source to source, the value 2.0 we used is consistent with many observations of dense clumps although the average value in molecular clouds is likely to be closer to $\sim$1.8  \citep{Beckwith1990, Boulanger1996, Hildebrand1983, PlanckCollaboration2011a, PlanckCollaboration2011b, PlanckCollaboration2011d}. Again,
the column density of hydrogen   is $N({\rm H_2})$ . From Eqs. \ref{eq:int} and \ref{eq:op_depth} we can derive the molecular column density
\begin{equation}
        N({\rm H_2}) \, = \, \frac{I_\nu}{B_\nu(T_{\rm dust})\kappa_\nu \mu}.
        \label{eq:dcd}
\end{equation}

The derived column densities from molecular lines and dust can be used to calculate the mass and density of a cloud or a clump as
\begin{equation}
\label{masseq}
        M_c \,  = \, N({\rm H_2}) \times \pi R^2 \times \frac{\mu}{M_\odot}
,\end{equation}
where $R$ is the radius of the clump. The average density, $n$, is calculated using the hydrogen column density
\begin{equation}
        n   =  \frac{N({\rm H_2})}{2R}
,\end{equation}
where the clump diameter $2R$ is the full-width at half maximum (FWHM) from the $Herschel$ column density map.


We derived the estimated virial mass  from the formula \mbox{\citep{MacLaren1988}}
\begin{equation}
         M_{\rm vir} \, = \, \frac{k\sigma^2 R}{G}
,\end{equation}
where $k$ depends on the density distribution. We choose $k=1.333, $ which corresponds to density distribution $\rho (r) \propto r^{-1.5}$. The velocity dispersion is given by the equation
\begin{equation}
         \sigma \, = \, \sqrt{\frac{kT_{kin}}{\overline m} + \left( \frac{\Delta V^2}{8ln(2)} - \frac{kT_{kin}}{m} \right)}
         \label{velocity_dispersion}
,\end{equation}
where $\overline m$ is the mean molecular mass (2.33~u assuming 10 \% He), and $m$ is the mass of the molecule used for observations. The temperature, $T_{kin}$, is the kinetic temperature, which is assumed to be 10~K. The velocity dispersion provides an estimate of the nonthermal motions that provide further support against gravity \citep{PlanckCollaboration2011c, Bertoldi1992}. If the cloud's mass is less than the virial mass, it is not gravitationally bound, and without external pressure for support, it will disperse.

We can also examine the stability using the model of BE spheres. The BE mass assumes a static isothermal cloud with no magnetic field \citep{McKee2007a}. We include the nonthermal component and use the effective sound speed instead of the isothermal sound speed. As a difference to the virial mass, the BE model includes the external pressure. The critical BE mass \citep{Bonnor1956} is
\begin{equation}
        M_{\rm BE} \, \approx \, 2.4 R_{\rm BE} \frac{\sigma^2}{G}
,\end{equation}
where $R_{\rm BE}$ is the BE radius, and $G$ is the gravitational constant.


\section{Results}
\label{results}

\subsection{Comparison of dust continuum and molecular line data}

First we compare the column densities calculated from the molecular line observations to the column densities calculated from the dust continuum observations. The sizes of the clumps were estimated as a FWHM from the dust column density maps and the physical sizes in pc are in Table \ref{colden}. Twelve of the clumps are larger than the size criteria (0.03 - 0.2 pc) given by \citet{Bergin2007} for cores, and, thus, are larger structures that could contain further substructures.

\begin{figure}
        \centering
        \includegraphics[scale = 0.49]{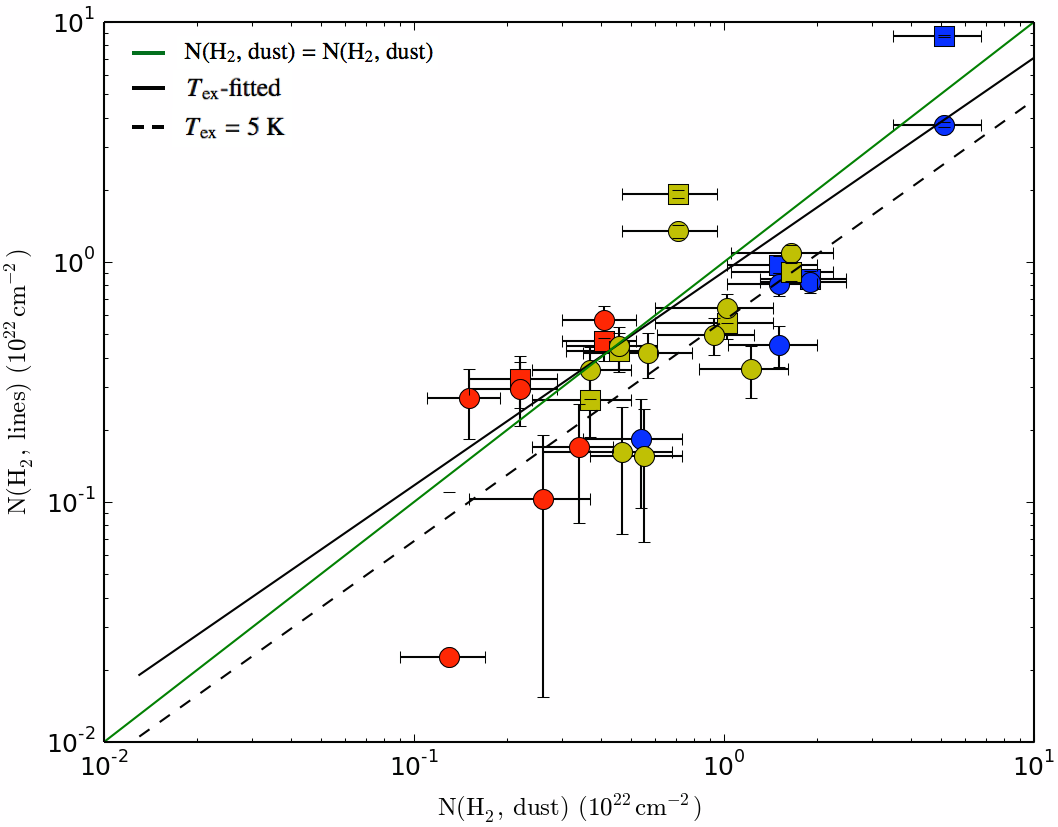}
        \caption [CO vs. dust]{Comparison of the column densities derived from molecular line and dust continuum observations. The squares are the column densities, where the excitation temperature could be calculated, and the circles are the column densities calculated with an assumed excitation temperature of 5~K. The colors indicate the dust color temperature. Blue is below 11~K, yellow is 11~-~13~K, and red is warmer than 13~K. The solid line is the linear least-squares fit to the column densities derived from the calculated excitation temperature, and the dashed line fits the $T_{\rm ex}$=5~K values. The green line indicates a line where $N(\rm H_2)_{\rm dust}$=$N(\rm H_2)_{\rm line}$.}
    \label{comp}
\end{figure}

We did not get a result for the excitation temperature in all of the cases. This is most likely because of the greater ratio between $^{13}$CO and C$^{18}$O abundances than the 5.5 that was assumed in the calculations. Thus, we initially used an estimate of $T_{\rm ex}$=10~K \citep[see, e.g.,][]{Dobashi1994, Alves1999}. The results from the calculations showed, however, that when $T_{\rm ex}$ could be estimated, it was typically 5~K (see Table \ref{colden}). Therefore, we show in Table \ref{colden} column densities for a fixed value of $T_{\rm ex}$=5~K.
 \begin{table*}
        \caption{Clump sizes (FWHM from dust column density), column densities derived from dust and line data, and estimated C$^{18}$O excitation temperatures.}
        \vspace{0,2cm}
        \centering
        \begin{tabular}{ |l|c|c|c|c|c|c| }
                \hline
                 & &  & $N($H$_2$) &  & $N($H$_2$)  & $N($H$_2$), $T_{\rm ex}$=5~K\\
        Field & FWHM ['] & FWHM [pc] & dust   [10$^{21}$ cm$^{-2}$]   & $T_{\rm ex}$ [K] & lines  [10$^{21}$ cm$^{-2}$]  & lines [10$^{21}$ cm$^{-2}$]\\

                \hline
                G86.97-4.06    & 0.4 & 0.17 & 12$\pm$4    &  -          &     -       & 4$\pm$1\\
                G92.04+3.93    & 0.8 & 0.38 & 50$\pm$20   & 4$\pm$2     & 90$\pm$40   & 37$\pm$8 \\
                G93.21+9.55    & 0.6 & 0.15 & 19$\pm$6    & 4.9$\pm$0.4 & 9$\pm$2     & 8$\pm$1\\
                G94.15+6.50    & 0.9 & 0.41 & 7$\pm$2     & 4$\pm$1     & 19$\pm$8    & 13$\pm$2\\
                G98.00+8.75    & 0.5 & 0.35 & 10$\pm$4    & 6$\pm$1     & 6$\pm$1     & 6.5$\pm$0.9\\
                G105.57+10.39  & 0.5 & 0.28 & 15$\pm$5    &  -          &     -       & 8$\pm$1\\
                G108.28+16.68  & 0.8 & 0.15 & 1.5$\pm$0.4 &  -          &     -       & 2.7$\pm$0.8\\
                G110.80+14.16  & 1.0 & 0.24 & 2.2$\pm$0.7 & 4.7$\pm$0.3 & 3.3$\pm$0.8 & 3.0$\pm$0.3\\
                G111.41-2.95   & 0.6 & 1.09 & 6$\pm$2     &  -          &     -       & 4.2$\pm$0.1\\ 
                G131.65+9.75 N & 0.6 & 0.38 & 9$\pm$3     &  -          &     -       & 5$\pm$1\\
                G131.65+9.75 S & 0.3 & 0.17 & 4$\pm$1     &  -          &     -       & 3.6$\pm$0.5\\
                G132.12+8.95   & 1.3 & 0.64 & 15$\pm$5    &  -          &     -       & 5$\pm$1\\
                G149.67+3.56   & 0.4 & 0.04 & 5$\pm$2     &  -          &     -       & 1.6$\pm$0.4\\
                G154.08+5.23   & 0.9 & 0.09 & 17$\pm$6    & 5.5$\pm$0.5 & 9$\pm$2     & 11$\pm$1\\
                G157.92-2.28   & 0.4 & 0.55 & 5$\pm$2     &  -          &     -       & 1.8$\pm$0.6\\
                G159.34+11.21  & 1.4 & 1.33 & 4$\pm$1     &  -          &     -       & 6$\pm$2\\
                G161.55-9.30   & 0.5 & 0.07 & 6$\pm$2     &  -          &     -       & 1.6$\pm$0.4\\
                G164.71-5.64   & 0.8 & 0.15 & 3$\pm$1     &  -          &     -       & 1.7$\pm$0.3\\
                G167.20-8.69   & 1.2 & 0.12 & 5$\pm$2     & 5.1$\pm$0.4 & 4$\pm$1     & 4.7$\pm$0.5\\ 
                G168.85-10.19  & 1.6 & 2.43 & 1.3$\pm$0.4 &  -          &     -       & 0.2$\pm$0.1\\
                G173.43-5.44   & 2.6 & 0.23 & 3$\pm$1     &  -          &     -       & 1.0$\pm$0.2\\
                \hline 
        \end{tabular}
        \label{colden}
\end{table*}
We plot the column densities calculated from molecular observations in comparison with the column density calculated from dust continuum observations in Fig. \ref{comp}.

In Fig. \ref{comp}, the cases where $T_{\rm ex}$ could be derived are plotted with square symbols, the solid line showing the corresponding linear least-squares fit. The column densities calculated with $T_{\rm ex}$=5~K are plotted with circles and fitted with a dashed line. In the cases where the excitation temperature could be calculated from the observations, the $T_{\rm ex}$=5~K values are usually within the margin of errors. The column densities calculated from dust are typically higher but also often within the margin of error. For example, for G159.34+11.21, for which $N(\rm H_2)_{\rm dust}$=4$\pm$1$\times10^{21}$~cm$^{-2}$ and $N(\rm H_2)_{5K}$=6$\pm$2$\times10^{21}$~cm$^{-2}$. In some cases the column densities are quite different, however. This is the case for  G86.97-4.06, where $N(\rm H_2)_{\rm dust}$=1.2$\pm$0.4$\times10^{22}$~cm$^{-2}$, while the value derived from the line data is much lower, $N(\rm H_2)_{5K}$=4$\pm$1$\times10^{21}$~cm$^{-2}$. Uncertainty of the background subtraction could contribute to some of these differences. For example, in G86.97-4.06 the reference region is estimated to have a column density of $\sim$10$^{21}$cm$^{-2}$, which  corresponds to a diffuse part of the cloud with little CO. However, the presence of a small column density of gas without CO molecules cannot explain the difference in the column density estimates of several times 10$^{21}$cm$^{-2}$.

The dust color temperatures of the clumps are shown in Fig. \ref{comp} in different colors: blue for clumps below 11~K, yellow for the range of 11$-$13~K, and red above 13~K. The column densities derived from dust show an anticorrelation with the dust color temperature as denser regions are cooler. This anticorrelation on the dust color temperature was not seen as clearly in the column densities derived from molecular line observations, as there are several column densities $\sim0.5\times10^{22}$~cm$^{-2}$ independent of the dust color temperature.

The maps of $T_{\rm mb}(^{13}{\rm CO})$ are shown in Figs. \ref{contour_maps1} and \ref{contour_maps2}. The contours correspond to the column densities derived from dust continuum data (subtracting the local diffuse background).
\begin{figure*}
    \begin{tabular}{cc}
    \centering
            \includegraphics[scale = 0.24]{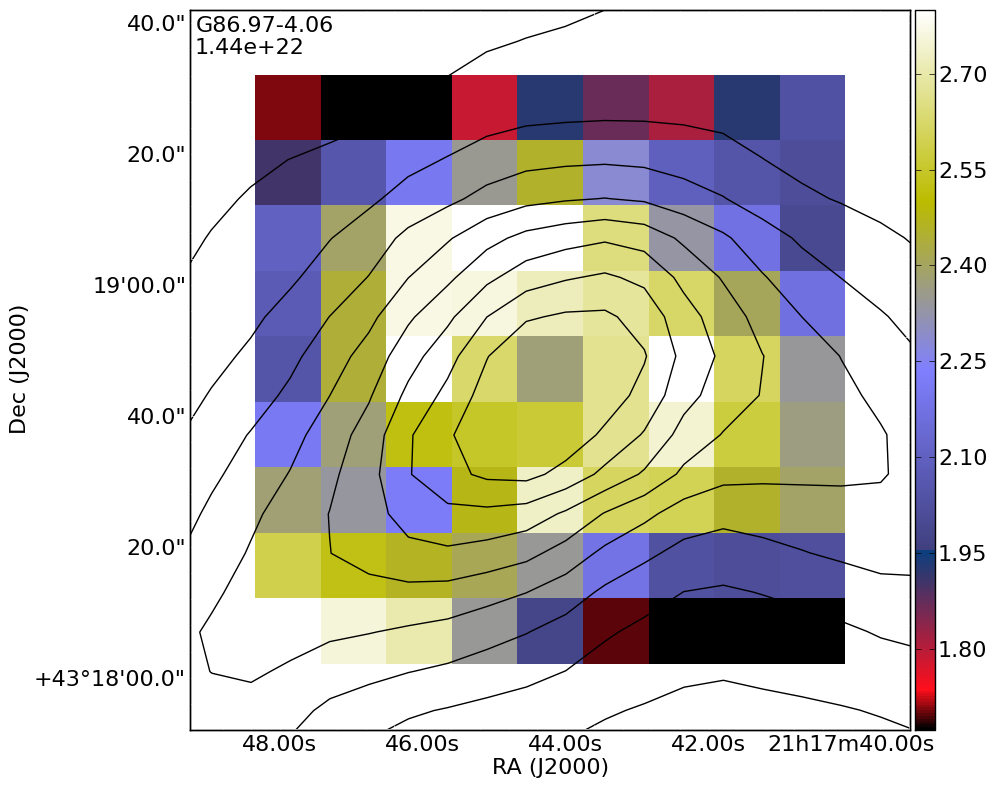}
        \includegraphics[scale = 0.24]{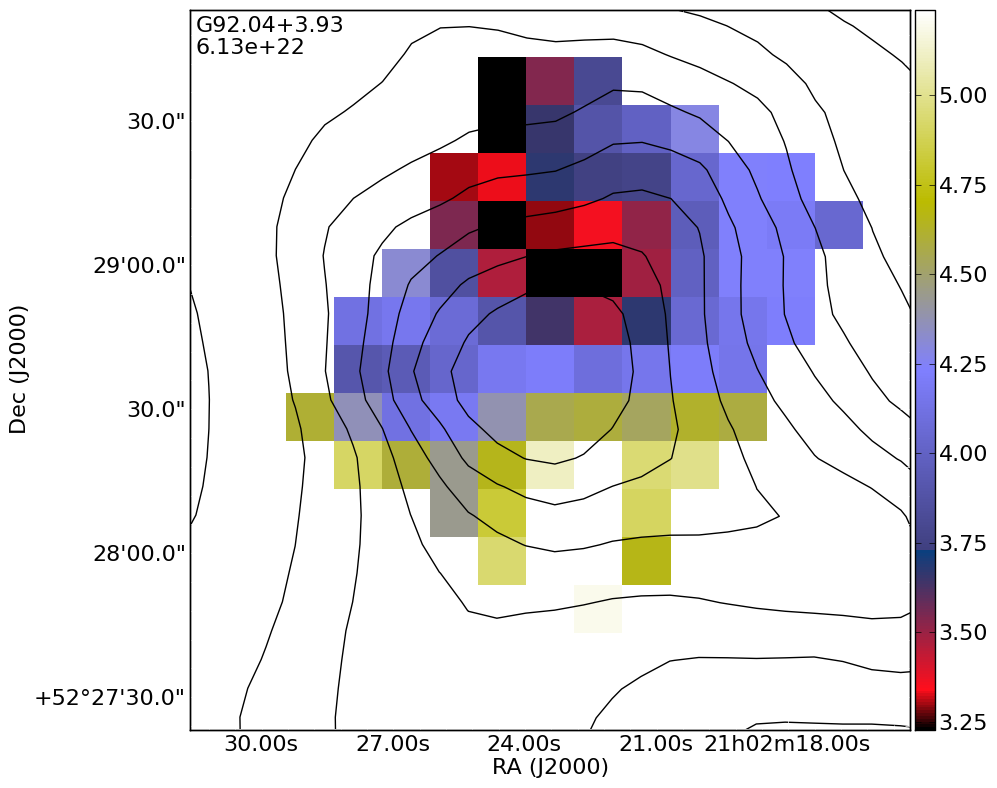}
            \includegraphics[scale = 0.24]{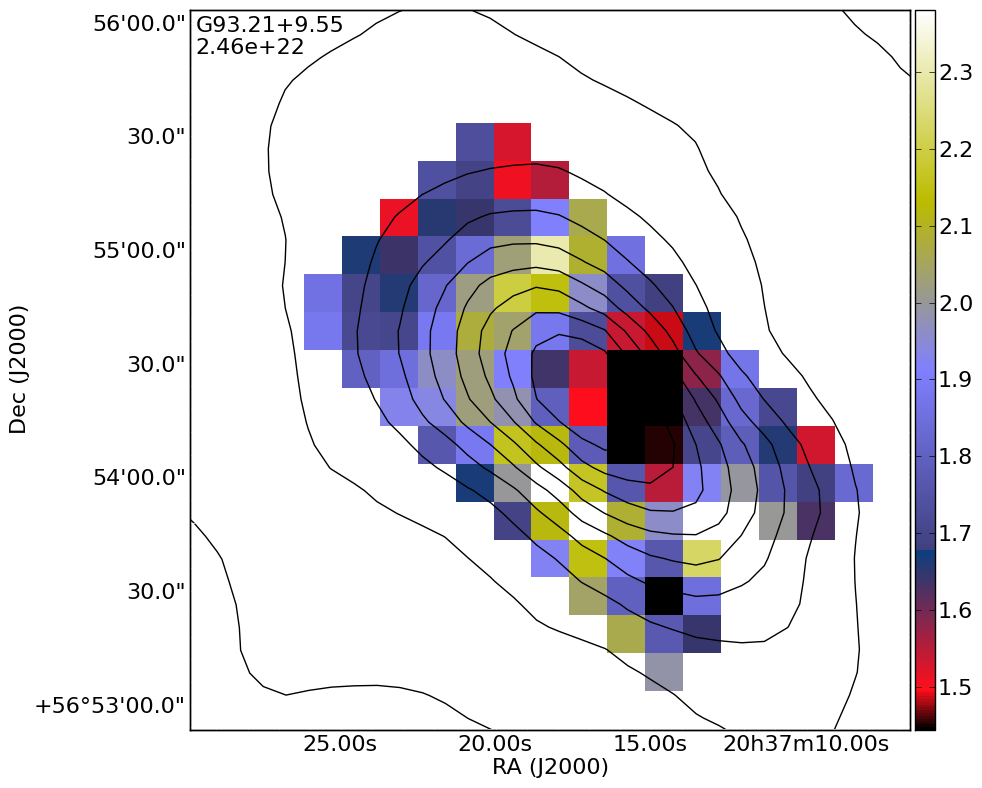}
        \end{tabular}
        
    \begin{tabular}{cc}
        \includegraphics[scale = 0.24]{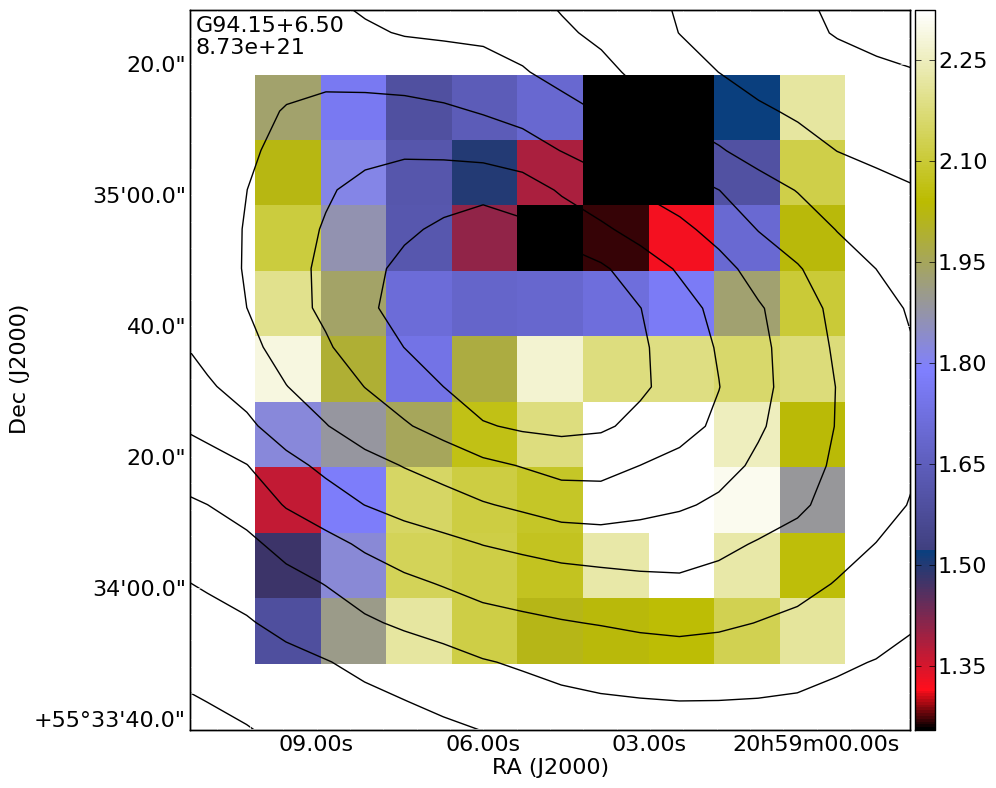}
            \includegraphics[scale = 0.24]{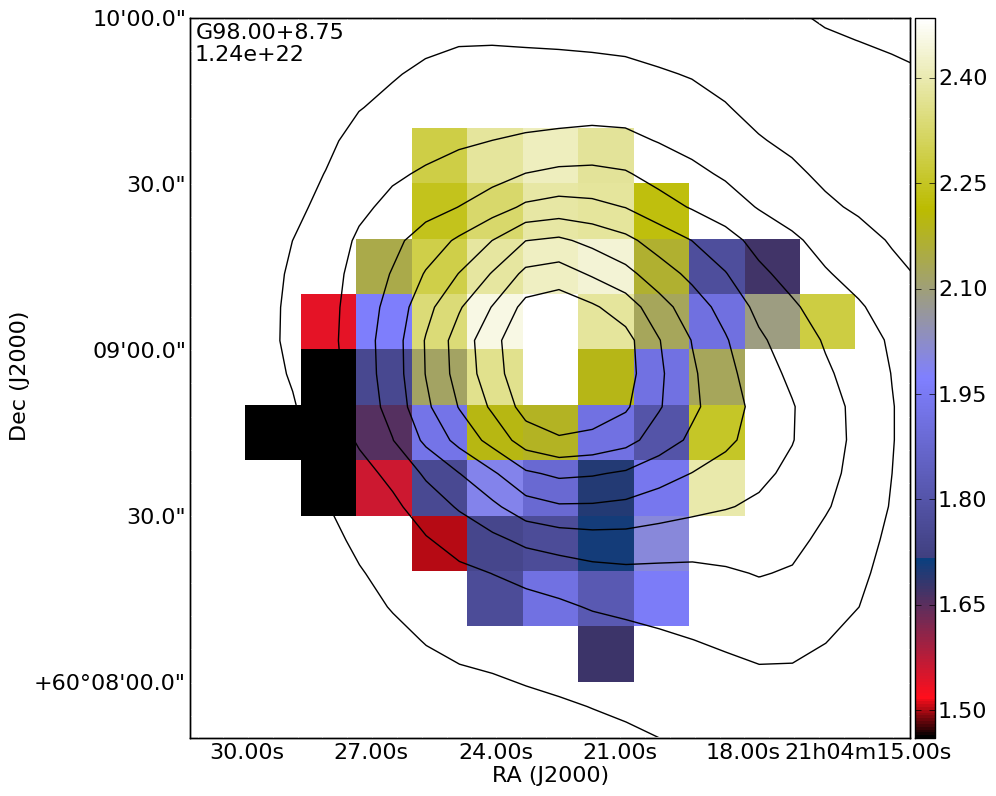}
            \includegraphics[scale = 0.24]{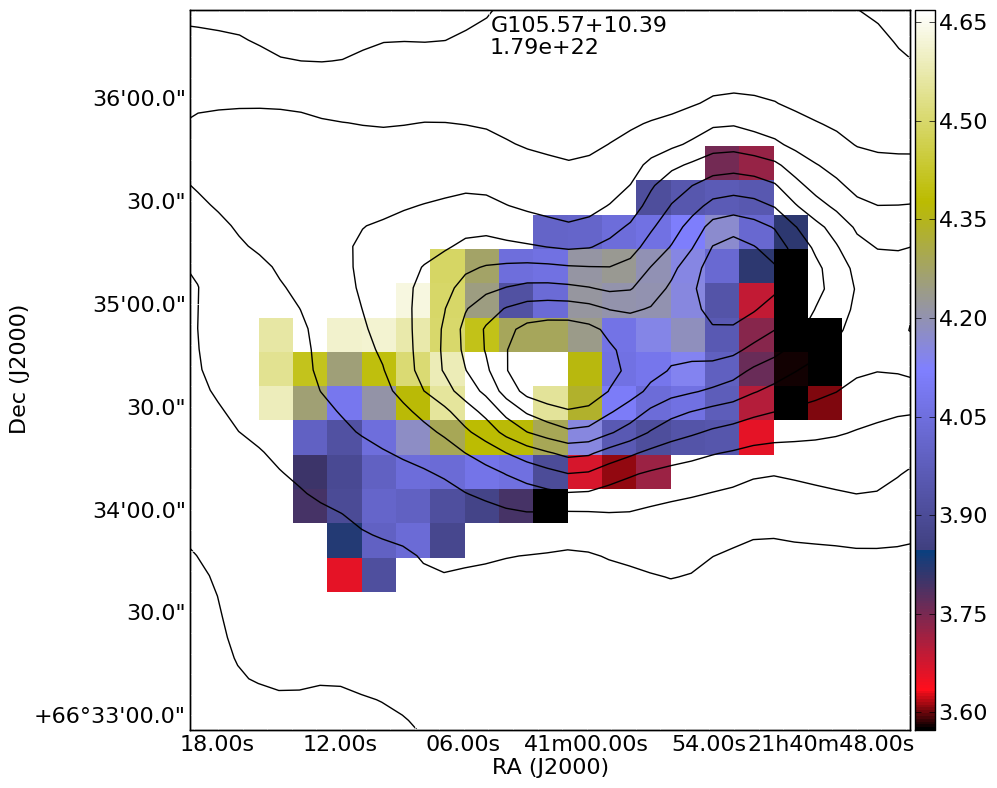}      
        \end{tabular}
        
        \begin{tabular}{cc}
            \includegraphics[scale = 0.24]{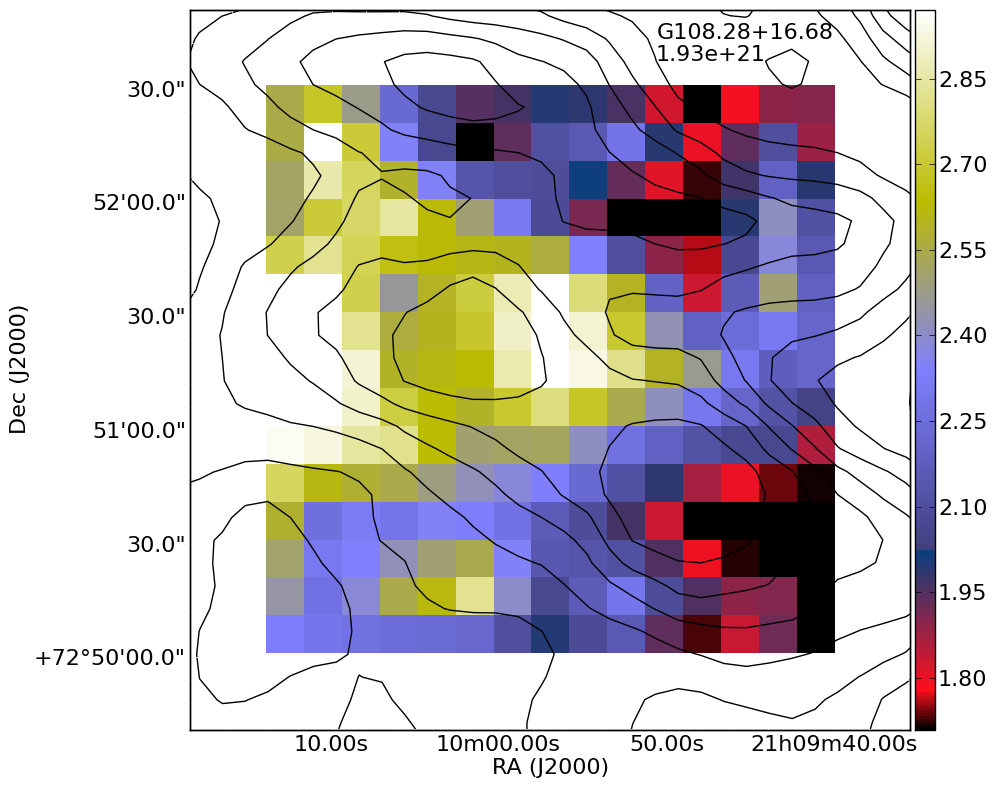}
        \includegraphics[scale = 0.24]{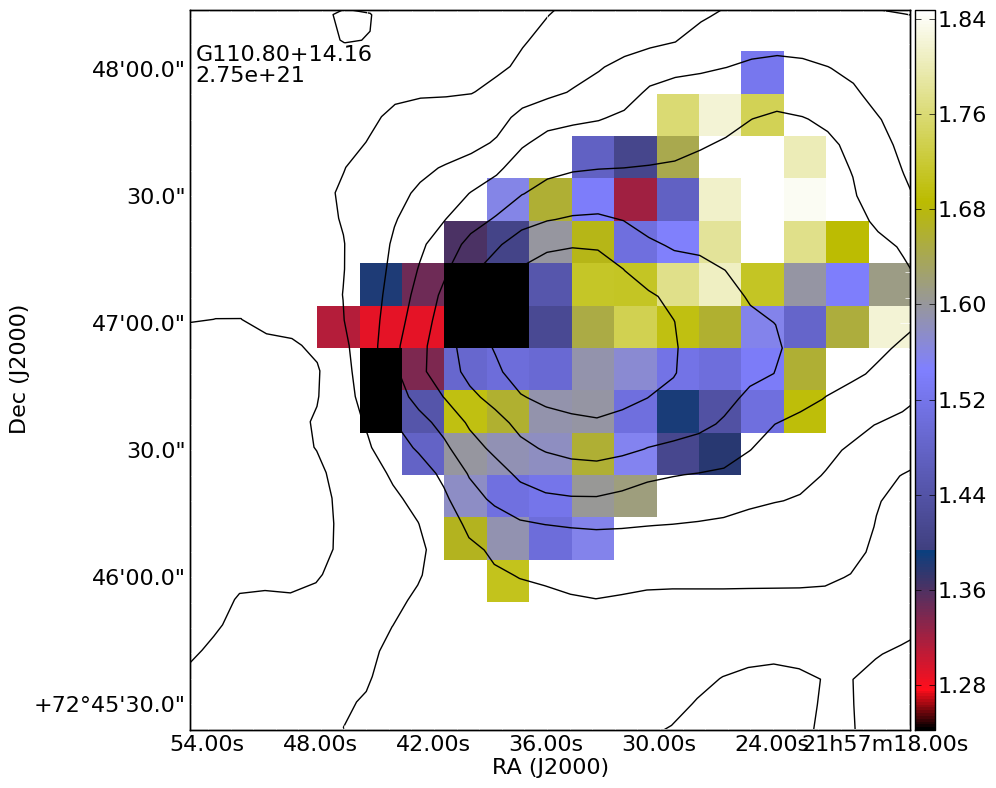}
            \includegraphics[scale = 0.24]{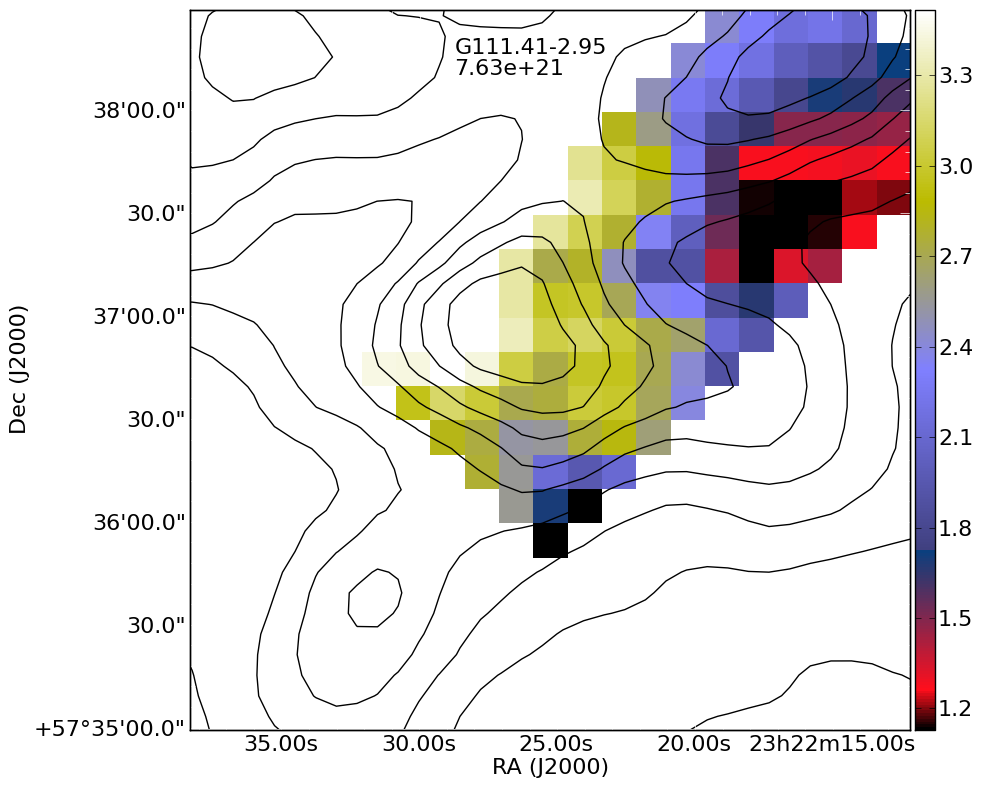}
        \end{tabular}

    \begin{tabular}{cc}
        \includegraphics[scale = 0.24]{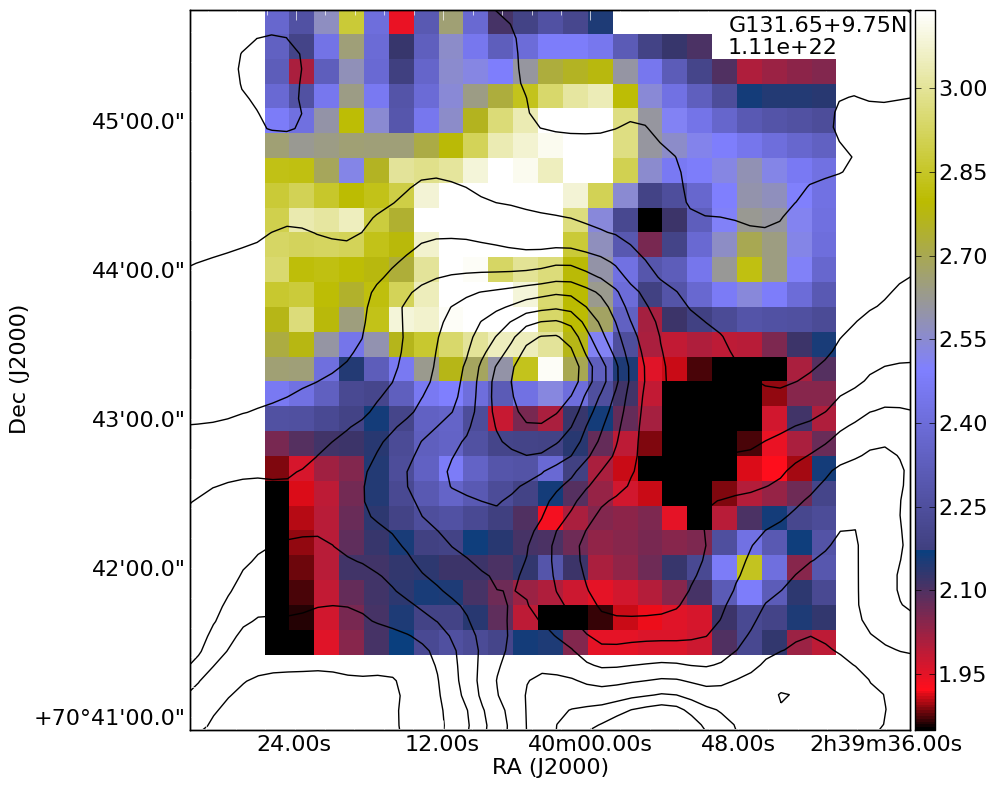}
            \includegraphics[scale = 0.24]{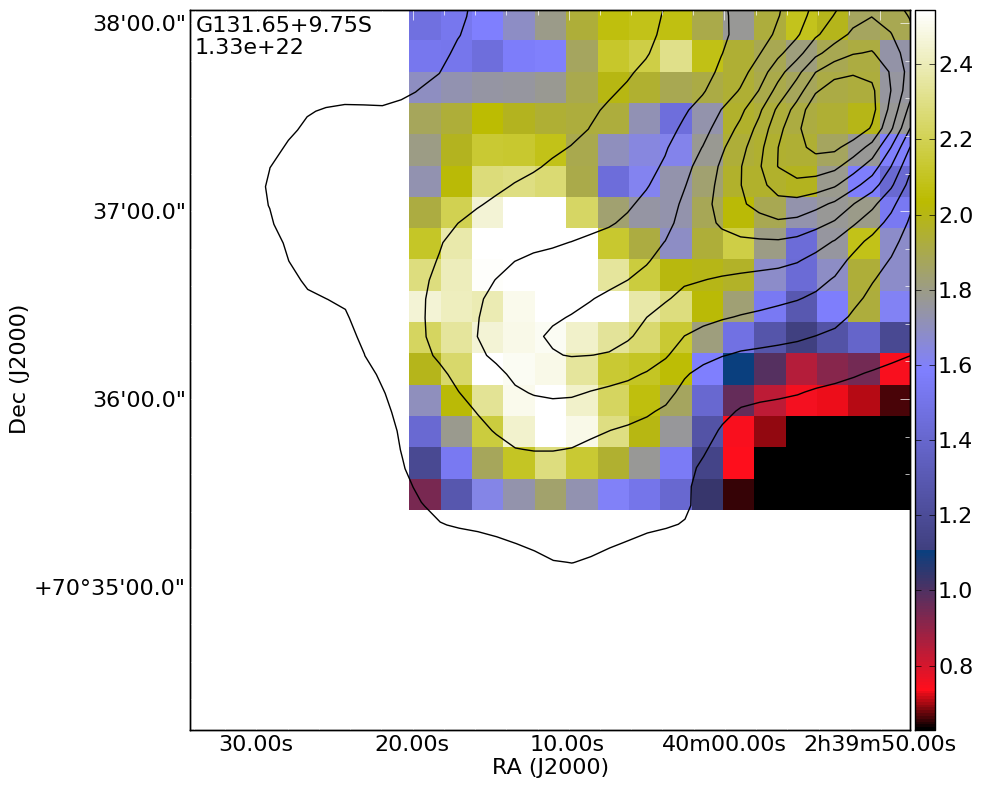}
            \includegraphics[scale = 0.24]{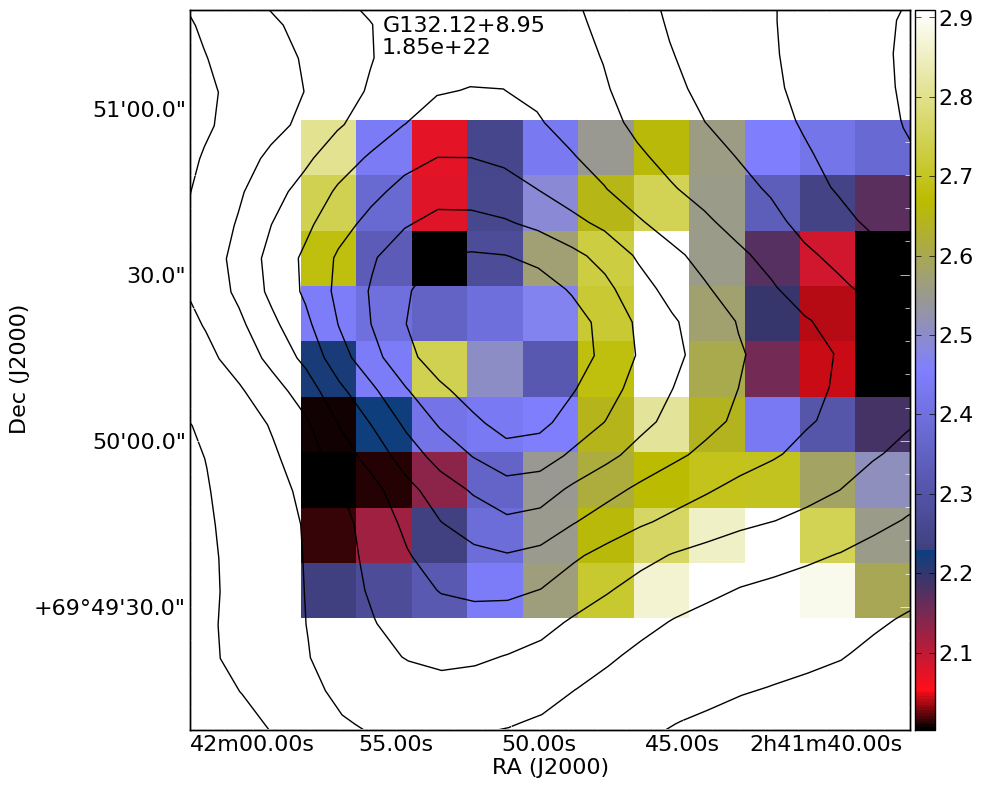}
        \end{tabular}
        \caption[Maps of $^{13}$CO main beam temperature]{Maps of $^{13}$CO main beam temperature ($T_{\rm mb}$ [K], resolution 35$\arcsec$). The contours show the column density derived from dust continuum observations (resolution 40$\arcsec$), with contour steps 10~\% of the peak value. The name of the field and the maximum value of N(H$_2$) derived from dust emission are marked in each frame.}
    \label{contour_maps1}
\end{figure*}   
\begin{figure*}
    \begin{tabular}{cc}
    \centering
            \includegraphics[scale = 0.24]{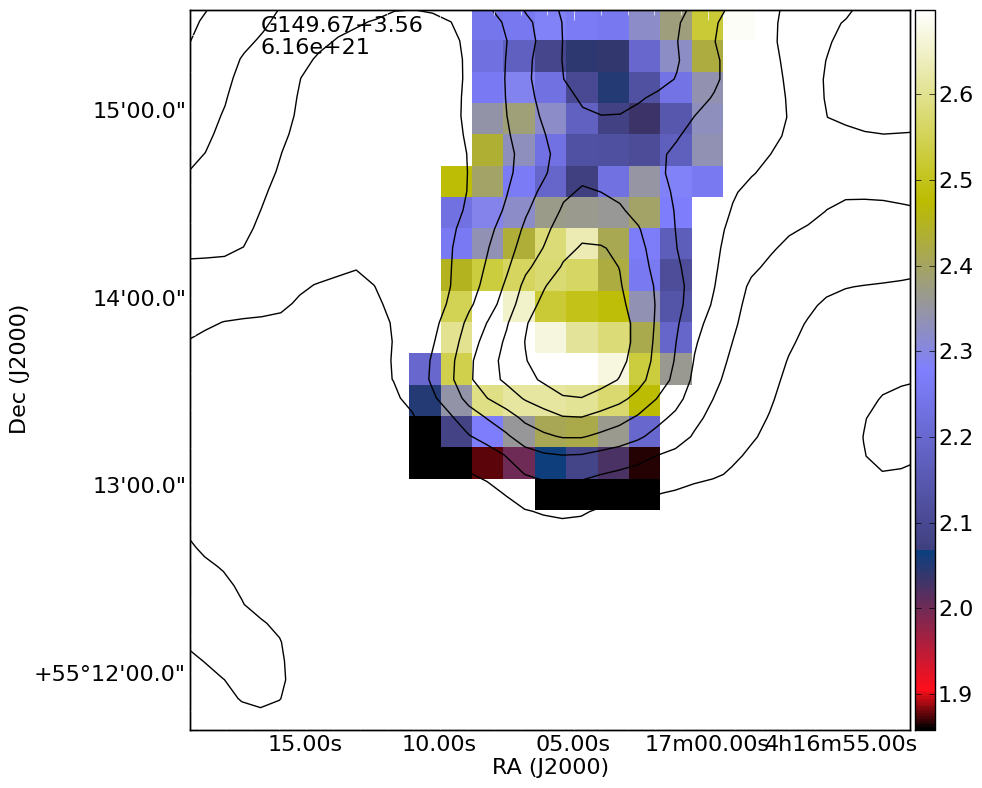}
            \includegraphics[scale = 0.24]{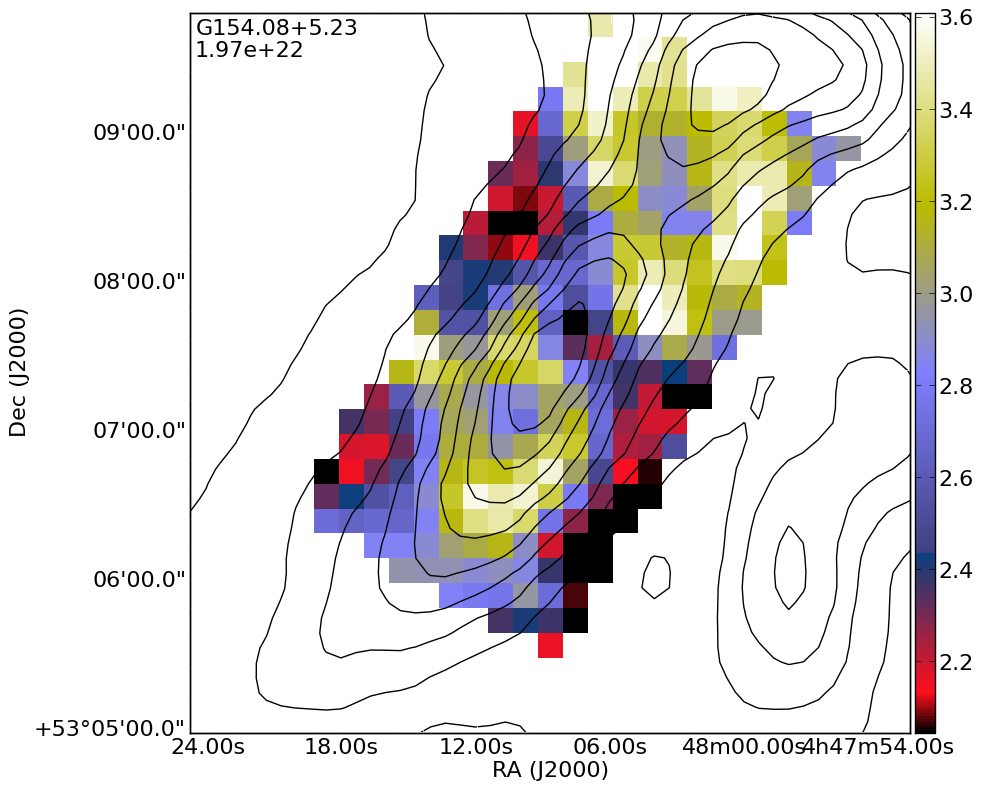}
            \includegraphics[scale = 0.24]{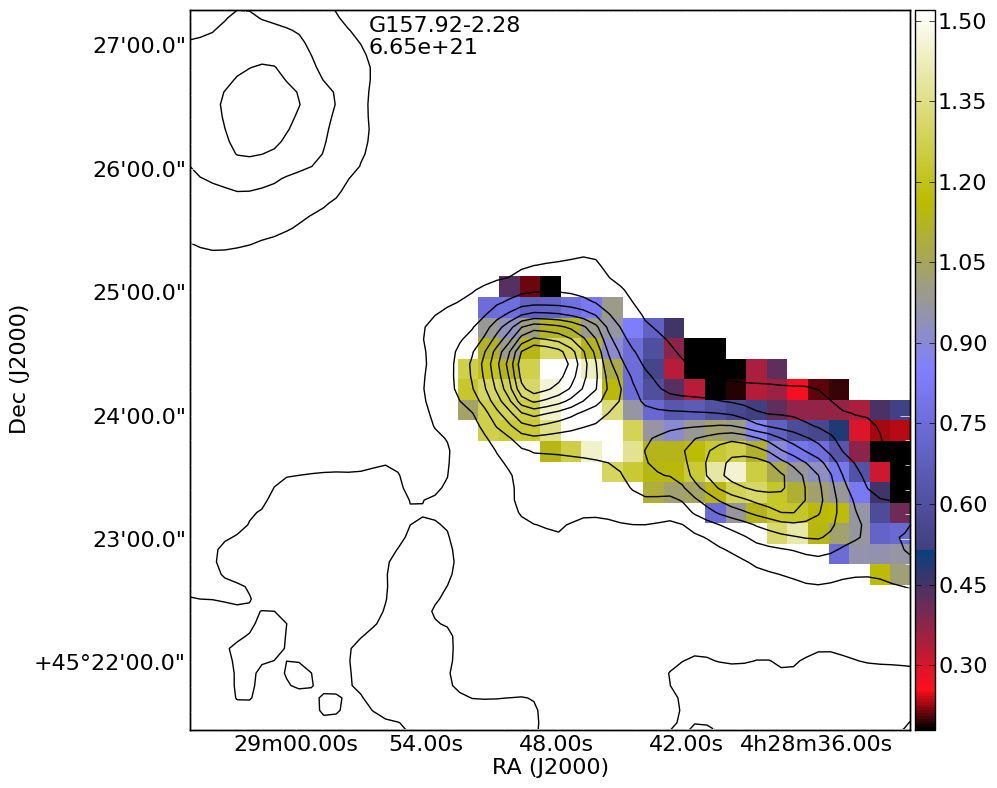}
        \end{tabular}
        \begin{tabular}{cc}
            \includegraphics[scale = 0.24]{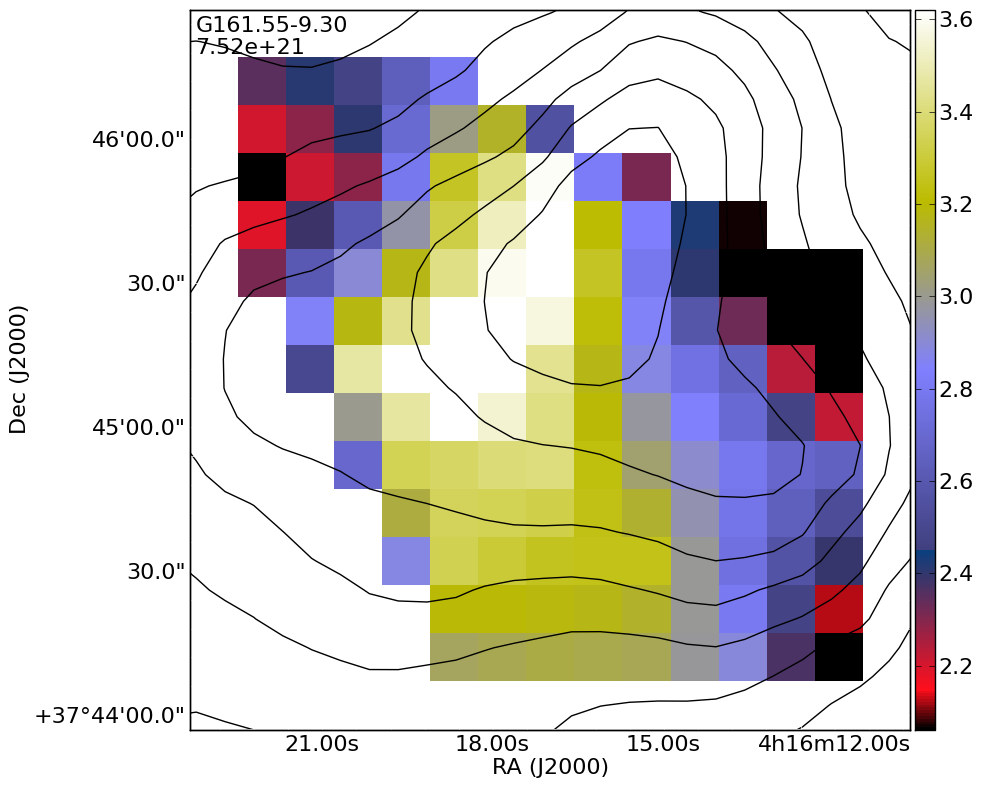}
            \includegraphics[scale = 0.24]{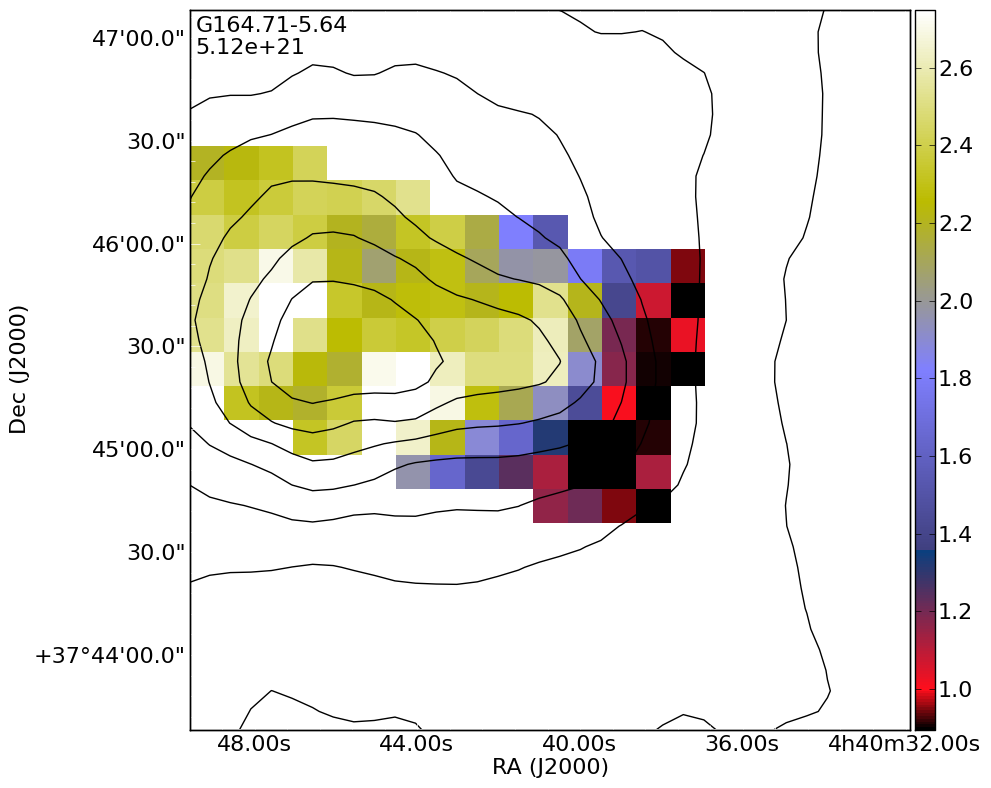}
            \includegraphics[scale = 0.24]{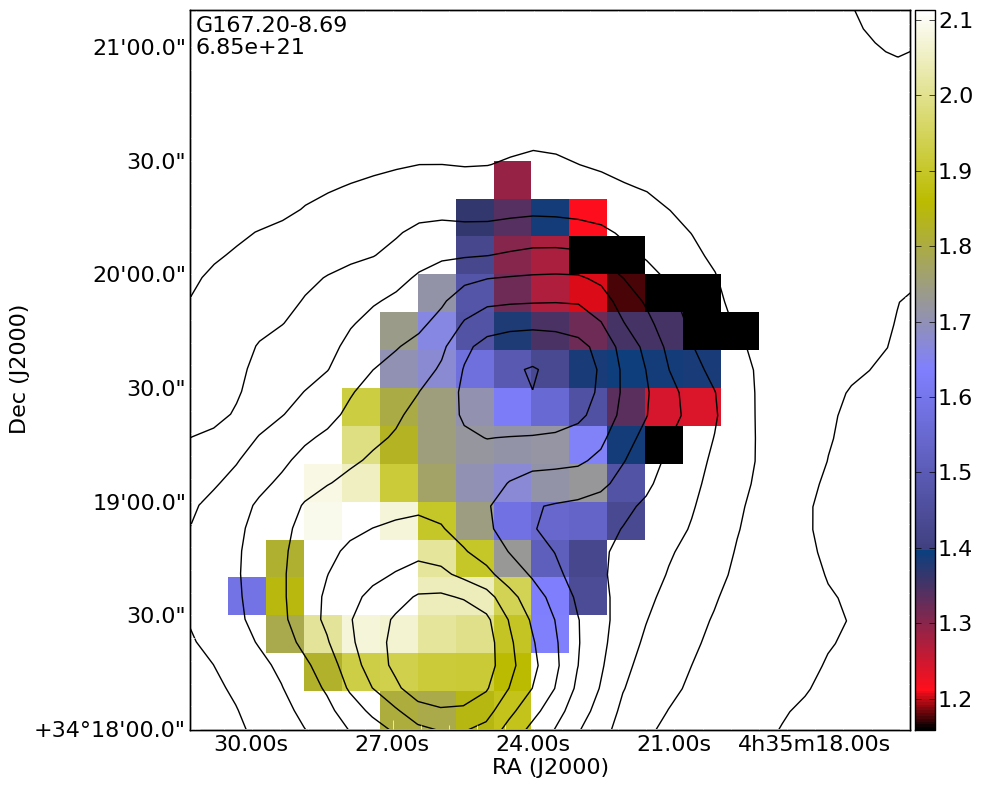}
    \end{tabular}
        \caption {Continued from Fig. \ref{contour_maps1}.}
    \label{contour_maps2}
\end{figure*}
The statistical error in the column density maps of the dust is at level 2$\times 10^{20}$~cm$^{-2}$. All the contours are at more than 10$\sigma$ level and, thus, statistically significant. The S/N of $^{13}$CO maps is lower. In a few cases, we get over 5$\sigma$ detection for the whole map, namely for fields G105.57+10.39 and G149.67+3.56. In five fields, G86.97-4.06, G108.28+16.68,
G131.65+9.75 S, G157.92.28, and G164.71-5.64,  the signal drops below 2$\sigma$ in part of the field.

In all the maps, we see a peak in the main beam temperature of $^{13}$CO close to the dust column density peak. For few fields, e.g., G98.00+8.75 (Fig. \ref{contour_maps1}), the peaks coincide, but for most cases there is a small shift, which can be up to 0.5$\arcmin$. In most cases the shift is smaller than the resolution of the observations. In some cases, as in G149.67+3.56, the peaks coincide partly, but are still shifted. The shifts are small in physical sizes as well, $\sim$0.01$-$0.22~pc, and less than the radius of the clumps.

Significant differences in the peak positions of dust and line emission are often observed because of molecular depletion, other chemical gradients, or because of temperature differences. \citet{Marsh2014} found shifts with a median distance of 59$\arcsec$ for H$^{13}$CO$^+$ and dust (250~$\muup$m). The dust and gas have been found to be coupled in the interior of cores with densities above 3$\times$10$^4$~cm$^{-3}$. At lower densities the temperature difference between the two components depends on the strength of the external UV field and not on density \citep{Galli2002}. Also, the depletion can increase the gas temperature in low and moderate densities \citep{Goldsmith2001}. If we compare the densities calculated from the molecular line observations in Table \ref{masses}, the clumps are less dense and, thus, dust and gas are not likely coupled in these clumps. However, the asymmetric displacement of the $^{13}$CO emission in the densest clumps is more likely to be a sign of variations of $^{13}$CO abundance.

\subsection{Clump masses}

The masses and densities along with the virial and BE masses are listed in Table \ref{masses}.
\begin{table*}
        \caption{The used line width of C$^{18}$O, densities, masses, virial masses, and BE masses for all the targets using molecular line observations and mass value calculated from the dust observations. The values from molecular line observations are derived with $T_{\rm ex}$ calculated from C$^{18}$O observations and with a fixed value of $T_{\rm ex}$~=~5~K.}
        \vspace{0,2cm}
        \centering
        \begin{tabular}{ |l|c|c|c|c|c|c|c|c| }
        \hline
         & $\Delta V_{\rm C^{18}O}$ &  & $n$ [10$^3$ cm$^{-3}$] &  & M$_{\rm line}$ [M$_{\odot}$] &  &  &   \\ 
    Field & [km s$^{-1}$] & $n$ [10$^3$ cm$^{-3}$] & ($T_{\rm ex}$=5~K) & M$_{\rm line}$ [M$_{\odot}$] & ($T_{\rm ex}$=5~K) & M$_{\rm dust}$ [M$_{\odot}$] & M$_{\rm vir}$ [M$_{\odot}$] & M$_{\rm BE}$ [M$_{\odot}$]  \\ 
    \hline 
    G86.97-4.06 & 0.74 & - & 7$\pm$2 & - & 1.7$\pm$0.5 & 6$\pm$2 & 10$\pm$2 & 6$\pm$2 \\ 
    G92.04+3.93 & 1.79 & 70$\pm$20 & 32$\pm$8 & 230$\pm$70 & 100$\pm$20 & 130$\pm$40 & 110$\pm$30 & 70$\pm$10 \\ 
    G93.21+9.55 & 0.65 & 19$\pm$4 & 18$\pm$2 & 3.2$\pm$0.7 & 3.1$\pm$0.4 & 7$\pm$2 & 7.5$\pm$0.8 & 4.5$\pm$0.5 \\ 
    G94.15+6.50 & 0.85 & 15$\pm$8 & 11$\pm$2 & 60$\pm$30 & 39$\pm$7 & 20$\pm$7 & 31$\pm$6 & 18$\pm$3 \\ 
    G98.00+8.75 & 0.76 & 5$\pm$1 & 6.0$\pm$0.8 & 12$\pm$2 & 13$\pm$2 & 22$\pm$9 & 22$\pm$3 & 13$\pm$2 \\ 
    G105.57+10.39 & 0.76 & - & 9$\pm$1 & - & 11$\pm$2 & 21$\pm$7 & 18$\pm$3 & 11$\pm$1 \\ 
    G108.28+16.68 & 1.07 & - & 6$\pm$2 & - & 1.0$\pm$0.3 & 0.6$\pm$0.2 & 16$\pm$4 & 10$\pm$2 \\ 
    G110.80+14.16 & 0.31 & 4.4$\pm$0.9 & 4.0$\pm$0.5 & 3.3$\pm$0.7 & 3.0$\pm$0.4 & 2.2$\pm$0.7 & 5.6$\pm$0.6 & 3.4$\pm$0.4 \\ 
    G111.41-2.95 & 2.27 & - & 1.2$\pm$0.3 & - & 90$\pm$20 & 120$\pm$50 & 500$\pm$100 & 290$\pm$80 \\ 
    G131.65+9.75 N & 1.22 & - & 4$\pm$1 & - & 13$\pm$3 & 23$\pm$8 & 50$\pm$10 & 30$\pm$8 \\ 
    G131.65+9.75 S & 0.53 & - & 7$\pm$1 & - & 1.7$\pm$0.2 & 1.8$\pm$0.6 & 6.5$\pm$0.9 & 3.9$\pm$0.6 \\ 
    G132.12+8.95 & 1.39 & - & 2.3$\pm$0.5 & - & 32$\pm$7 & 110$\pm$30 & 110$\pm$30 & 70$\pm$20 \\ 
    G149.67+3.56 & 0.64 & - & 12$\pm$3 & - & 0.05$\pm$0.01 & 0.15$\pm$0.07 & 2.1$\pm$0.5 & 1.3$\pm$0.3 \\ 
    G154.08+5.23 & 0.56 & 33$\pm$7 & 40$\pm$4 & 1.3$\pm$0.3 & 1.5$\pm$0.2 & 2.3$\pm$0.8 & 3.7$\pm$0.5 & 2.2$\pm$0.3 \\ 
    G157.92-2.28 & 0.87 & - & 1.1$\pm$0.3 & - & 9$\pm$3 & 30$\pm$10 & 40$\pm$10 & 26$\pm$8 \\ 
    G159.34+11.21 & 0.84 & - & 1.4$\pm$0.4 & - & 180$\pm$60 & 130$\pm$30 & 100$\pm$30 & 60$\pm$20 \\ 
    G161.55-9.30 & 0.48 & - & 7$\pm$2 & - & 0.13$\pm$0.04 & 0.4$\pm$0.1 & 2.3$\pm$0.7 & 1.4$\pm$0.4 \\ 
    G164.71-5.64 & 0.50 & - & 3.6$\pm$0.7 & - & 0.7$\pm$0.1 & 1.4$\pm$0.4 & 6$\pm$1 & 3.4$\pm$0.7 \\ 
    G167.20-8.69 & 0.32 & 12$\pm$2 & 12$\pm$1 & 1.0$\pm$0.2 & 1.1$\pm$0.1 & 0.47$\pm$0.05 & 2.8$\pm$0.3 & 1.7$\pm$0.2 \\ 
    G168.85-10.19 & 0.21 & - & 0.03$\pm$0.01 & - & 20$\pm$10 & 130$\pm$40 & 50$\pm$30 & 30$\pm$10 \\ 
    G173.43-5.44 & 0.24 & - & 1.4$\pm$0.2 & - & 0.9$\pm$0.1 & 2$\pm$1 & 4.5$\pm$0.7 & 2.7$\pm$0.4 \\ 
        \hline 
        \end{tabular}
        \label{masses}
\end{table*}
The uncertainties for densities and masses were determined with the Monte Carlo method. The uncertainties are big in many cases, up to 53~\%, and we must be careful when drawing conclusions from the density and masses. The mass errors do not include the error for distance ($\sim$30~\%, in the case of G168.85-10.19 the error is closer to $\sim$50~\%), which further increases the uncertainty. In addition, in most cases the excitation temperature could not be calculated and the excitation temperature of 5~K was assumed. In the cases where the excitation temperatures could be calculated, the derived densities and masses are within the margin of error of those derived from the dust emission data. The field G92.04+3.93 is the only exception.

The calculated masses vary greatly between the clumps, from one tenth of a solar mass to a couple of hundred. Thus, they clearly cover a mix of different structural scales. Following the mass criteria of \citet{Bergin2007}, 11 of them are more massive than typical cores ($> ocal5 {\rm M_{\odot}}$). However, most of them show a single kinematic component and are therefore consistent with the idea of clumps as velocity coherent objects. If both M$_{\rm line}$ and M$_{\rm dust}$ are larger than the virial and BE mass, we consider them to be gravitationally bound, and thus (potentially) on the verge of collapse. If M$_{\rm line}$ or M$_{\rm dust}$ exceeds BE mass, the object is potentially bound, but if the estimated mass is smaller than both the virial and BE mass, we consider the object not to be prestellar.

Two clumps, G92.04+3.93 and G159.34+11.21, appear to be on the verge of collapse. The mass of G92.04+3.93 calculated with $T_{\rm ex}$=5~K is lower than the virial mass, but when excitation temperature could be calculated it exceeds the virial mass clearly. For the field G159.34+11.21, the excitation temperature could not be calculated, thus, the calculated mass is based on the assumption of $T_{\rm ex}$=5~K. Three other clumps, G94.15+6.50, G98.00+8.75, and G105.57+10.39, are supercritical, albeit not significantly, if we consider the BE mass. The remaining 16 clumps seem to be near equilibrium.

\subsection{Observations of ${N_2H^{+}}$}

To estimate the ${\rm N_2H^{+}}$ column densities, we performed fits of the hyperfine structure of the J=1-0 lines. The fits were carried out assuming a single excitation temperature. We treated the excitation temperature, optical depth of the 23-12 component, and the radial velocity and width of the spectral lines  as free parameters \citep[see][]{Pirogov2003}. Because of the low signal-to-noise ratio, the optical depths and excitation temperatures were uncertain. Therefore we calculated the column density  assuming a fixed value of the excitation temperature, $T_{\rm ex}$=5~K, and optically thin emission \citep{Caselli2002a}. The line parameters and obtained column densities are listed in Table~\ref{n2h+_results}. We estimated the error estimates  with the Monte Carlo method, however, excluding the effect of the uncertainty of $T_{\rm ex}$. Apart from G86.97-4.06, shown in Fig.~\ref{n2h+_g86}, the ${\rm N_2H^{+}}$ spectra are shown in Appendix~\ref{N2H+_app}.

\begin{table*}
        \caption{The main beam temperature, velocity, noise and column density of N$_2$H$^+$ in the fields it was detected.}
        \vspace{0,2cm}
        \centering
        \begin{tabular}{ |l|c|c|c|c|c|}
        \hline
        Field & $T_{\rm mb}$ [K] & V [km s$^{-1}$] & $\Delta$V [km s$^{-1}$] & rms [K]  & N [10$^{12}$ cm$^{-2}$]  \\
        \hline 
        G86.97-4.06   &  0.34$\pm$0.03 &   5.50$\pm$0.02 &  0.43$\pm$0.04 & 0.06 &  1.1$\pm$0.3 \\ 
        G92.04+3.93   &  0.17$\pm$0.01 &  -1.80$\pm$0.02 &  0.51$\pm$0.04 & 0.04 & 3.0$\pm$0.2 \\  
        G93.21+9.55   &  0.23$\pm$0.04 &  -1.73$\pm$0.03 &  0.41$\pm$0.08 & 0.07 &  2.1$\pm$0.5 \\  
        G98.00+8.75   &  0.23$\pm$0.02 &   4.98$\pm$0.03 &  0.54$\pm$0.04 & 0.06 & 2.3$\pm$0.6 \\ 
        G105.57+10.39 &  0.31$\pm$0.02 & -10.10$\pm$0.02 &  0.46$\pm$0.04 & 0.05 & 2.0$\pm$0.5 \\  
        G132.12+8.95  &  0.32$\pm$0.02 & -12.41$\pm$0.02 &  0.51$\pm$0.04 & 0.04 & 1.1$\pm$0.3 \\
        \hline 
        \end{tabular}
        \label{n2h+_results}
\end{table*}

\begin{figure}
    \centering
        \includegraphics[scale = 0.8]{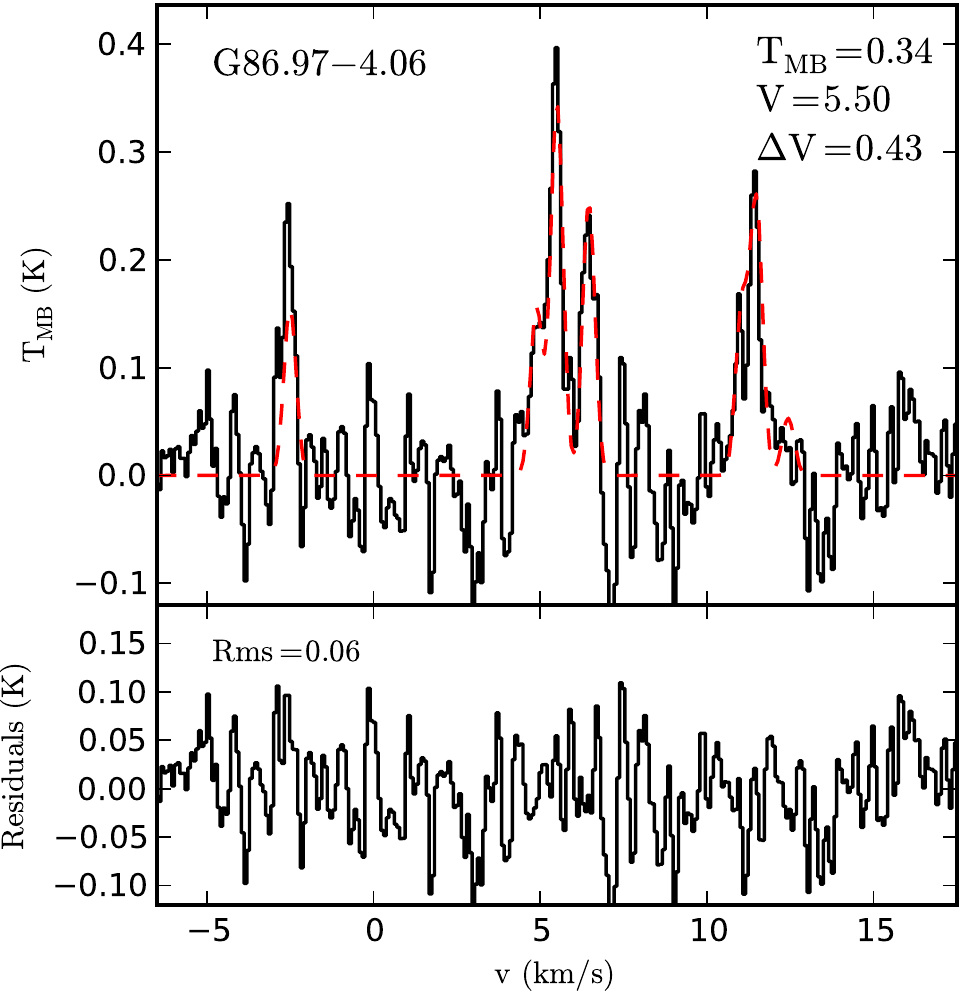}

    \caption[]{The ${\rm N_2H^{+}}$ spectrum observed toward G86.97-4.06. The red dashed line is the fit to the hyperfine spectra and the lower subframe shows the fit residuals. The main beam temperature of the 23-12 component, the fitted radial velocity, the FWHM line width (km~s$^{-1}$), and the residual rms noise (as main beam temperature) are given in the frames.} 
    \label{n2h+_g86}
\end{figure}

For most of the studied clumps, N$_2$H$^+$ line was not detected, suggesting that these clumps are not dense enough for significant CO depletion and the associated rise of N$_2$H$^+$ abundance \citep{Bergin1997, Charnley1997a}. The line was detected in six clumps: G86.97-4.06, G92.04+3.93, G93.21+9.55, G98.00+8.75, G105.57+10.39, and G132.12+8.95 (see Fig. \ref{n2h+_g86} for an example of the spectra and Table \ref{n2h+_results}). These clumps were among those with coldest dust temperatures. The main line in the fields was $T_{\rm mb} \sim$0.2$-$0.8 K and the satellites are barely visible above the noise of $\sim$0.05~K. The low excitation temperature of the CO lines can also be an indication of a low density, as in lower densities the molecules are less thermalized and the excitation temperature is much lower than the kinetic temperature. In the studied clumps, the excitation temperature of the C$^{18}$O in most cases was $\sim$5~K indicating a clear difference between the excitation and kinetic temperatures.

The calculated ${\rm N_2H^{+}}$ column densities are only a few times 10$^{12}$~cm$^{-2}$. These can be compared with the ${\rm N_2H^{+}}$ survey of nearby, low-mass cores conducted by \citet{Caselli2002a}. In that study, for the low-mass cores, $M_{\rm vir}\sim 10 $~M$_{\sun}$, the average ${\rm N_2H^{+}}$ column density was $\sim$7$\times 10^{12}$~cm$^{-2}$, and they found starless cores had a marginally lower column density than protostellar cores. Our column densities correspond to a lower end of those values. Conversely, \citet{Pirogov2003} observed a sample of 35 massive molecular cloud cores (mean virial mass over 600 solar masses) and found an average column density of 29$\times 10^{12}$~cm$^{-2}$. Our low column density values are consistent with the low mass of the clumps, but also suggest that the densities (and CO depletion) are not significant enough for a large ${\rm N_2H^{+}}$ abundance.
     
\subsection{Modeling}
        
The observations were compared to radiative transfer models of CO lines and of dust continuum emission. The modeling is explained in more detail in Appendix \ref{rtm} and we present  only the main results. The modeling was limited to the two positions in the field G131.65+9.75. These positions were chosen as they were in the same field and, thus, enable comparison of clumps at the cloud edge and in its center.

The calculations that are based on the dust model of \citet{Ossenkopf1994} find $^{13}$CO abundances that are close to the value of $10^{-6}$. Because the abundances depend greatly on the column density estimated from continuum data, the values are uncertain. The abundance ratio for $^{13}$CO and C$^{18}$O was found to be $\sim$10, instead of the 5.5 assumed in Sect. \ref{methods}, independent of the different assumptions used in the modeling (see Table \ref{table:modeling}). \citet{Harjunpaeae1996} and \citet{Anderson1999} have also found similar abundance ratios in dark clouds and cores.


\section{Discussion}
\label{discussion}

In about one third of the clumps, where excitation temperature could be estimated, the column densities are similar to those derived from dust (see Fig. \ref{comp} and Table \ref{colden}). The column densities calculated with $T_{\rm ex}$=10~K, were lower by a factor of three. Using $T_{\rm ex}$=5~K, which was more in line with those that could be calculated, we get very similar results, especially with the cases where column densities are lower ($\sim$10$^{21}$~cm$^{-2}$). Higher column densities tend to have larger uncertainties. For example, the column density for clump G92.04+3.93, 5$\pm$2$\times$10$^{22}$~cm$^{-2}$ from dust and 9$\pm$4$\times$10$^{22}$~cm$^{-2}$ from molecular lines is still inside the margin of errors. The cases where $T_{\rm ex}$=5~K were assumed to have mostly lower column densities than those where $T_{\rm ex}$ was solved. One explanation for this could be that CO has been depleted in these areas and, thus, the column density and the volume density calculated from CO observations are too low. The most likely explanation for the difference to the dust estimates is, however, the excitation temperature that is likely to vary from clump to clump.

\begin{figure}
    \centering
        \includegraphics[scale = 0.45]{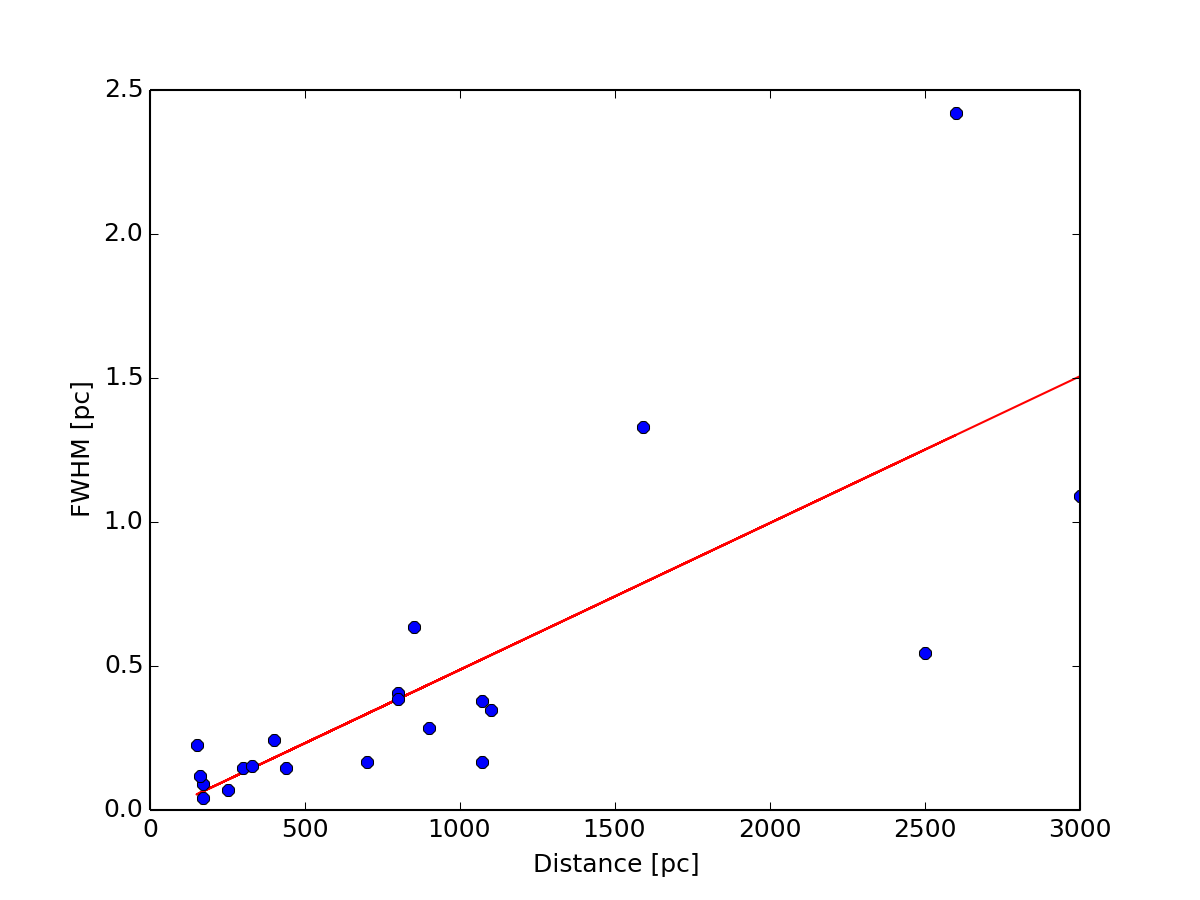}
        \caption {FWHM (pc) vs. distance with the best fit.}
    \label{dist_vs_size}
\end{figure}

The distances vary from 150~pc to 3~kpc across the sample, but the physical sizes (see Table \ref{colden} and Fig. \ref{dist_vs_size}) of the objects are, with one exception, $\sim$1~pc or below and thus comparable to typical clump scales as defined by \citet{Bergin2007}.
The largest sizes include those of the most distant clouds, namely, G111.80+14.16, G157.92-2.28, G159.34+11.21, and G168.85-10.19. On average, the physical size is proportional to 0.77 times the distance, a natural result of our resolution at larger distances corresponding to larger structures.

Although there are large uncertainties in the column densities derived from the lines, even with the calculated $T_{\rm ex}$ values, the mass and column density estimates indicate that the selected clumps might not be very dense, and subsequently may not be actual prestellar cores. The lack of N$_2$H$^+$ emission also supports this conclusion, with upper limits $T_{\rm mb}$(N$_2$H$^+$)=0.05$-$0.2~K for the 16 of 21 clumps.

The fact that the excitation temperature could not be calculated in all of the selected positions could be caused by the abundance ratio of $^{13}$CO and C$^{18}$O , which was actually higher than the assumed upper limit of 5.5. When extinction A$_V$ is small, the abundance ratio can range much above the terrestrial value \citep{Minchin1995}. The abundance ratio we found in our models was $\sim$10, compared to the used terrestrial value of 5.5. We found the abundance for $^{13}$CO, however,  to be close to the canonical 10$^{-6}$, provided that we trust the dust-derived column densities. The difference in the estimated abundance ratio can be caused by a real differences in the abundances or excitation temperatures or by optical depth effects. The modeling, however, roughly takes  the effect of optical depth and the expected difference in the excitation temperature of the isotopomers into account. This difference is caused by the difference in the photon trapping resulting from the difference in the optical depth of the lines. The assumption of the homogeneous source in the LTE analysis is inaccurate and, in reality, the excitation temperature of C$^{18}$O is probably below the excitation temperature of $^{13}$CO (and not equal as assumed) and drops precipitously toward the cloud surface. If the $^{13}$CO is optically thick, the observed intensity originates in a different part of the cloud than the C$^{18}$O. Our calculations did not show $^{13}$CO to be optically thick, but we calculated the optical thickness using the assumption that the excitation temperatures are equal. Because of the smaller optical depth, more of the C$^{18}$O line photons escape, leading to a smaller excitation temperature. This is also clearly significant when the kinetic temperature varies within the cloud \citep{Juvela2012b}. 

Based on the line data, 16 of the studied clumps are subcritical and, even considering the uncertainties of the mass estimates, only five sources could be gravitationally bound. In three cases, G92.04+3.93,
G94.15+6.50, and G159.34+11.21, the calculated mass is larger than the virial and BE mass even when considering the uncertainty. Two other clumps, G98.00+8.75 and G105.57+10.39, are within error margins when we consider the BE mass. The five clumps are from  the entire temperature range. However, only one clump, namely G92.04+3.93, had a mass larger than the virial and BE masses on a level $>1\sigma$, although G92.04+3.93 does not exceed this level if we assume $T_{\rm ex}$=5~K. If we consider the error in the distance of the object, none of the sources exceed the \mbox{$1\sigma$-level}. However, only four clumps have masses below the virial and BE masses on a level $>1\sigma$. In the coldest supercritical clumps, G92.04+3.93 ($T_{\rm dust}=$10.4~K), G98.00+8.75 ($T_{\rm dust}=$11.6~K), and G105.57+10.39 ($T_{\rm dust}=$10.9~K), we detect the N$_2$H$^+$ line, which would support the idea that these are being denser and possibly prestellar. The N$_2$H$^+$ emission seemed in general to be connected to the coldest clumps of $\sim$10$-$11~K, although it was not observable in all the clumps of these temperatures. For the clumps where N$_2$H$^+$ was detected, we get a FWHM line width of $\lesssim$0.5~km s$^{-1}$, values typically found in prestellar cores \citep{Johnstone2010}. 

If we rely on the mass estimates derived from dust observations instead of CO lines, the results are similar, only three clumps are below the virial mass and BE mass limits. Four of the clumps were larger by more than the error margins and could be classified as prestellar. The four clumps were G92.04+3.93, G105.57+10.39, G159.34+11.21, and G168.85-10.19. Thus, most of the clumps that were detected for low dust color temperature appear to disperse with time rather than form stars. \citet{Meng2013} and \citet{Wu2012} had similar results in their cold cores surveys: only a small fraction of the clumps were found to be prestellar. Of the 673 sources, 10 clumps were mapped and 22 potential cold cores were identified by \citet{Wu2012}. Of these 22 cores, seven were found to be gravitationally bound. Our results are in general agreement with these results on the fractions of starless and prestellar cores (5 out of 21 were gravitationally bound). \citet{Meng2013} studied  a subset of \citet{Wu2012} in more detail, and 90 \% of them were found to be starless (4 cores out of a sample of 38 were associated with sources). Studies of protostellar sources should be extended to a larger set of Planck clumps, including the 21 clumps that we discuss.


\section{Conclusions and summary}
\label{conclusions}

We investigated 21~clumps, that show low dust color temperatures T$\sim$10$-$15~K and are therefore potential places of star formation. To study the conditions in the clouds, we used molecular line and dust continuum observations. Our comparison of the two tracers shows that, even though the gas and dust emission are mainly morphologically compatible, the dust peak is sometimes shifted up to 30$\arcsec$ relative to the $^{13}$CO maximum. The data were examined with standard LTE analysis and with radiative transfer modeling to understand better the physical and chemical state of the clumps.

The column density calculations from molecular lines were mostly within the uncertainties to those calculated from dust, when the excitation temperature could be calculated. With an assumed fixed excitation temperature value of $T_{\rm ex}$=5~K, the derived column densities were only slightly below the dust estimates.

Using modeling to compare dust continuum and line data, the abundance of $^{13}$CO was found to be close to the typically assumed value of 10$^{-6}$. However, the abundance ratio of $^{13}$CO and C$^{18}$O was $\sim$10, higher than the terrestrial value 5.5.

The calculated masses had big uncertainties and one must be careful when drawing conclusions. When the masses were compared to virial and BE masses, only five clumps had a high enough mass to be gravitationally bound. In three of these clumps, we also found N$_2$H$^+$, so the clumps found in the fields G92.04+3.93, G98.00+8.75, and G105.57+10.39 could be prestellar. G92.04+3.93 and G105.57+10.3 were also among the coldest  $T_{\rm dust}\sim$10-11~K, while G98.00+8.75 was slightly warmer $T_{\rm dust}=$12~K. The highest temperatures in the examined clumps were $T_{\rm dust}\gtrsim$13~K.

The stability of the clumps requires further study. Observations of higher CO isotopomer transitions and further density and temperature tracers are needed to  better quantify the gas component in these cold clumps.

\appendix

\section{$^{13}$CO and C$^{18}$O spectra}
\label{co_spectra_app}

A sample spectrum of $^{13}$CO of all the fields are shown in Fig. \ref{spectra_app_pic1} and \ref{spectra_app_pic2}. The spectra are from the $^{13}$CO peak position of each field. The C$^{18}$O spectrum is shown from the same position; the baseline has been moved to -2.0~K.

\begin{figure*}
    \centering
        \includegraphics[scale = 0.29]{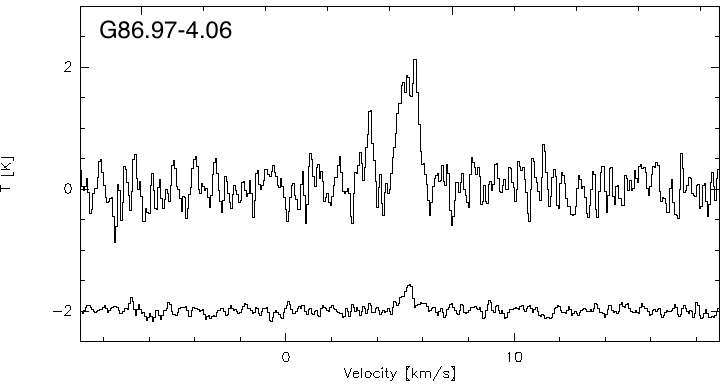}
        \includegraphics[scale = 0.29]{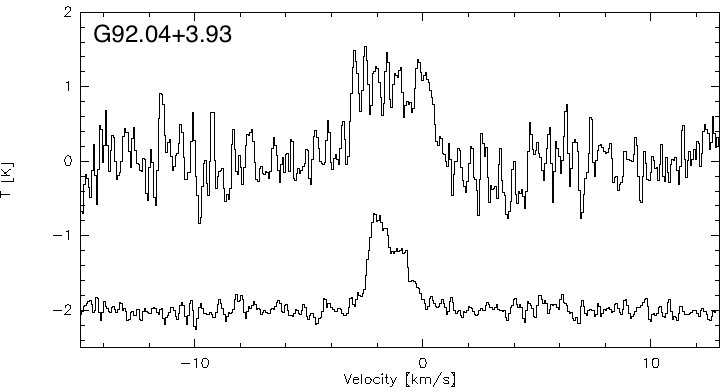}
        \includegraphics[scale = 0.29]{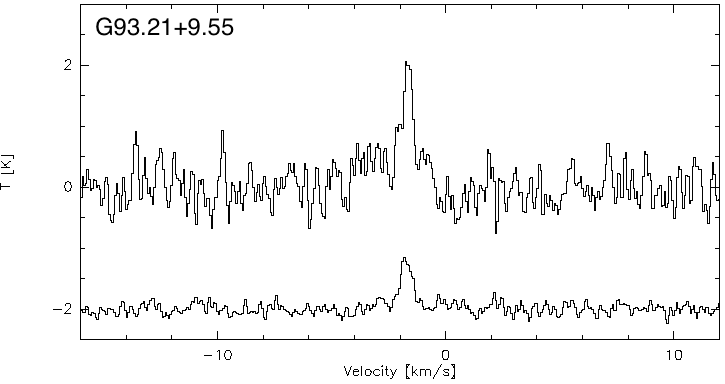}
        \includegraphics[scale = 0.29]{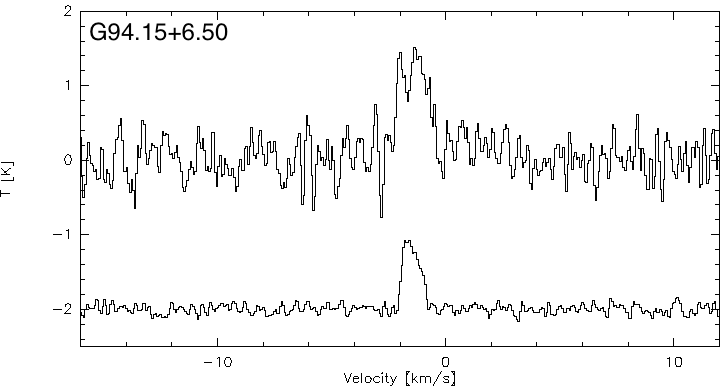}
        \includegraphics[scale = 0.29]{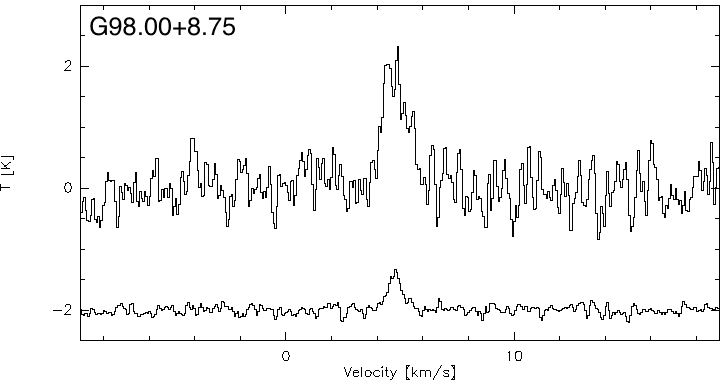}
        \includegraphics[scale = 0.29]{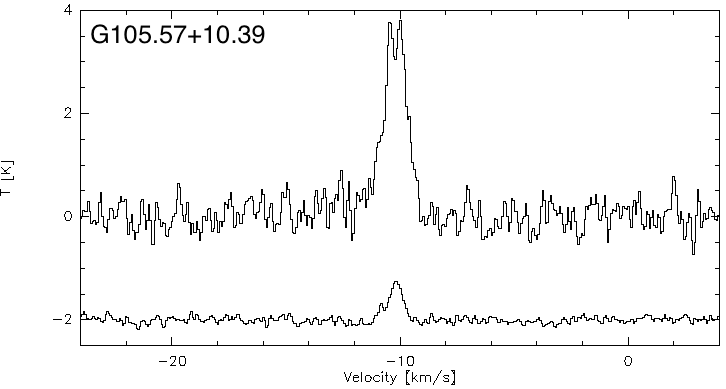}
        \includegraphics[scale = 0.29]{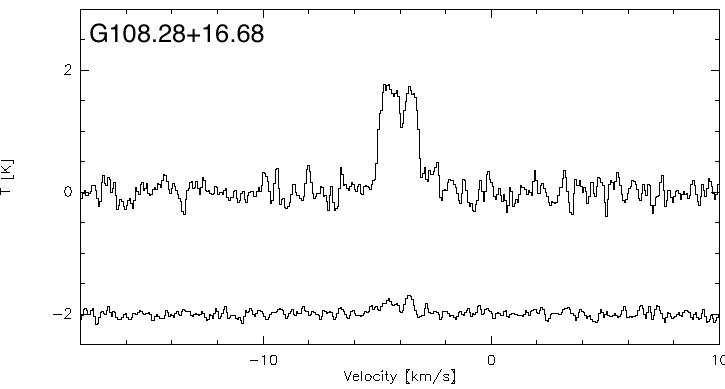}
        \includegraphics[scale = 0.29]{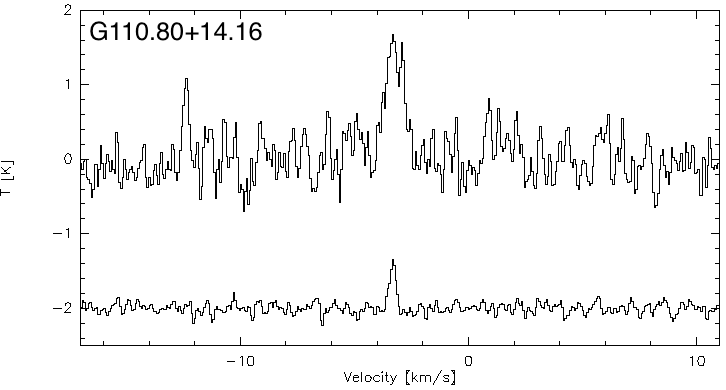}
        \includegraphics[scale = 0.29]{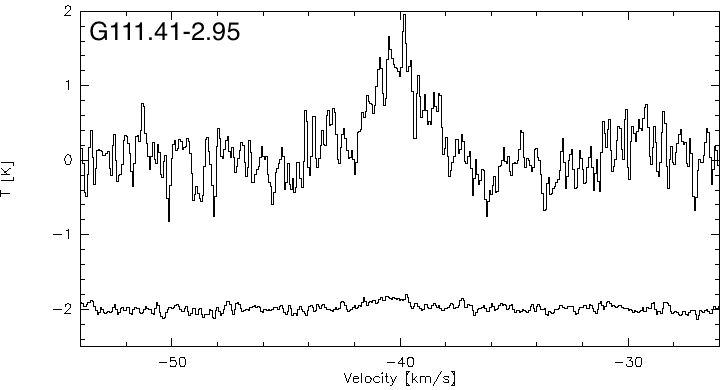}
        \includegraphics[scale = 0.29]{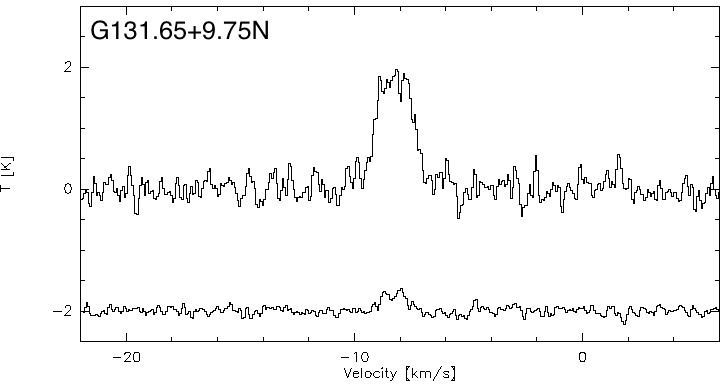}
        \includegraphics[scale = 0.29]{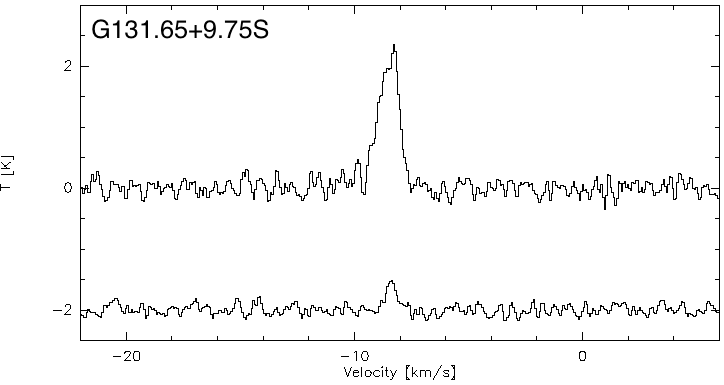}
        \includegraphics[scale = 0.29]{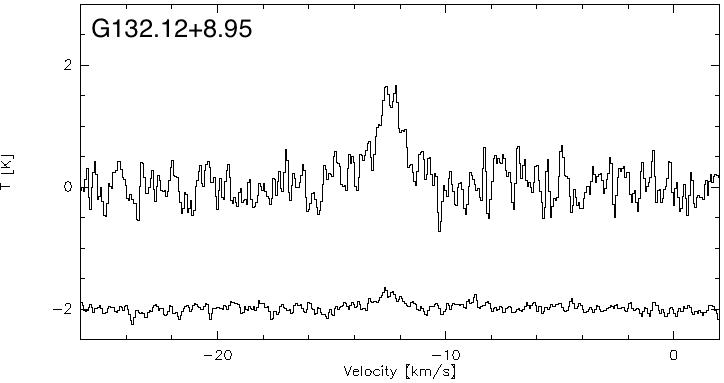}
    \caption[]{The observed $^{13}$CO and C$^{18}$O spectra at the $^{13}$CO peak position. The x-axis shows the velocity and the y-axis shows the main beam temperature. For plotting, the C$^{18}$O spectra have been shifted by 2~K.}
    \label{spectra_app_pic1}
\end{figure*}

\begin{figure*}
    \centering
        \includegraphics[scale = 0.29]{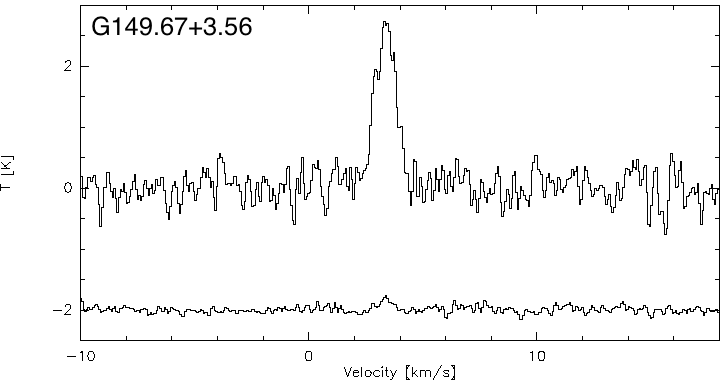}
        \includegraphics[scale = 0.29]{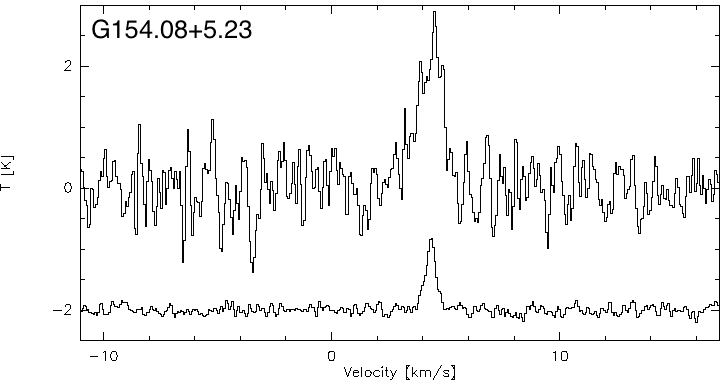}
        \includegraphics[scale = 0.29]{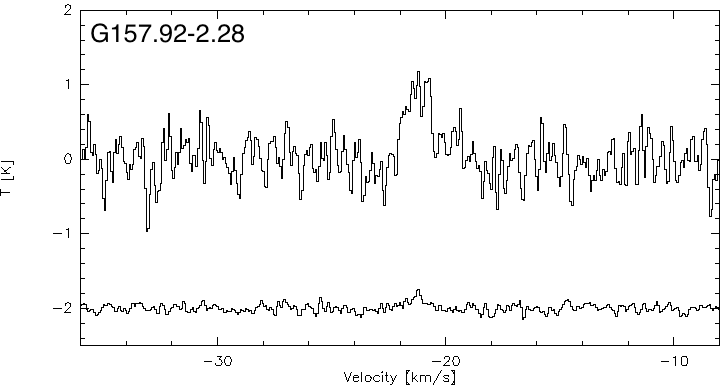}
        \includegraphics[scale = 0.29]{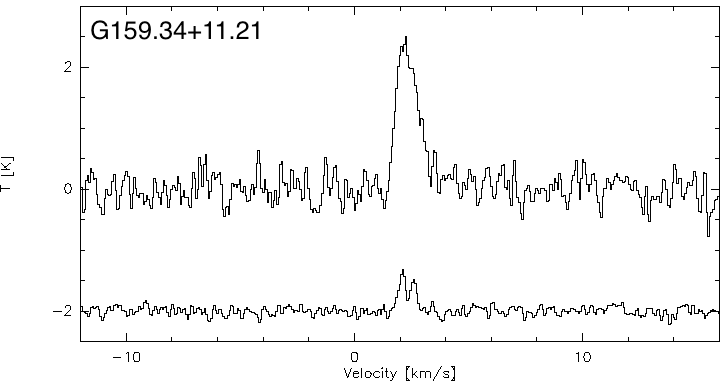}
        \includegraphics[scale = 0.29]{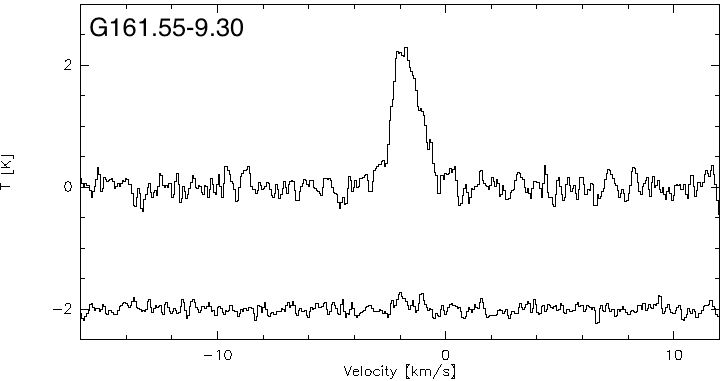}
        \includegraphics[scale = 0.29]{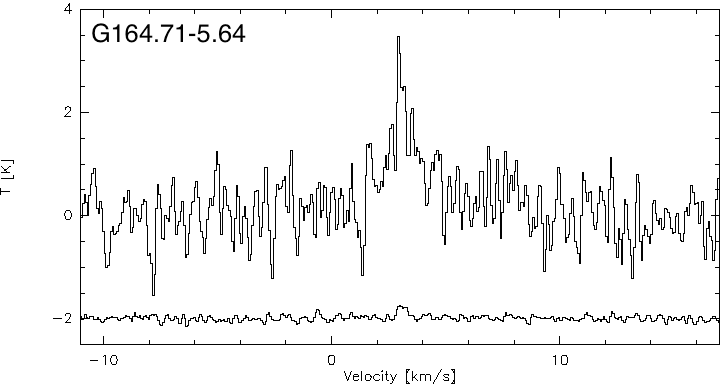}
        \includegraphics[scale = 0.29]{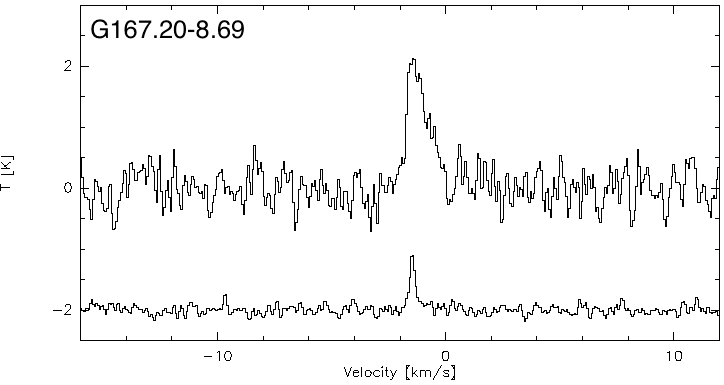}
        \includegraphics[scale = 0.29]{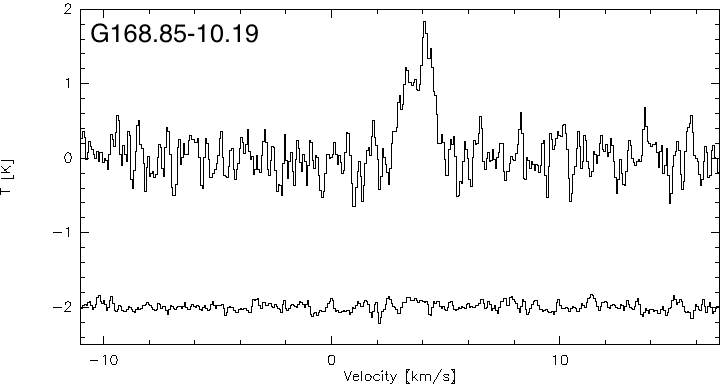}
        \includegraphics[scale = 0.29]{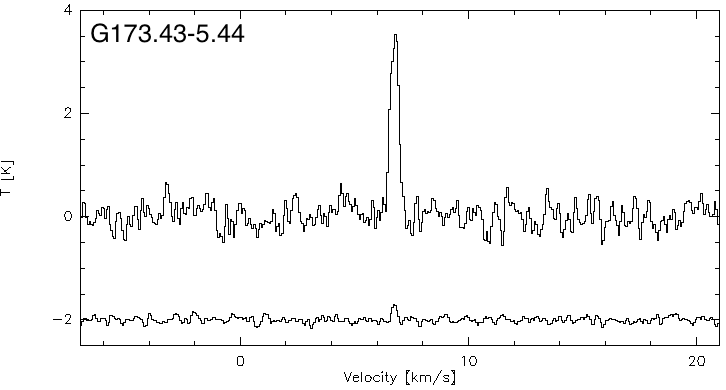}
    \caption[]{Continued from Fig. \ref{spectra_app_pic1}.}
    \label{spectra_app_pic2}
\end{figure*}

\section{N$_2$H$^+$}
\label{N2H+_app}

\begin{figure*}
    \centering
        \includegraphics[scale = 0.69]{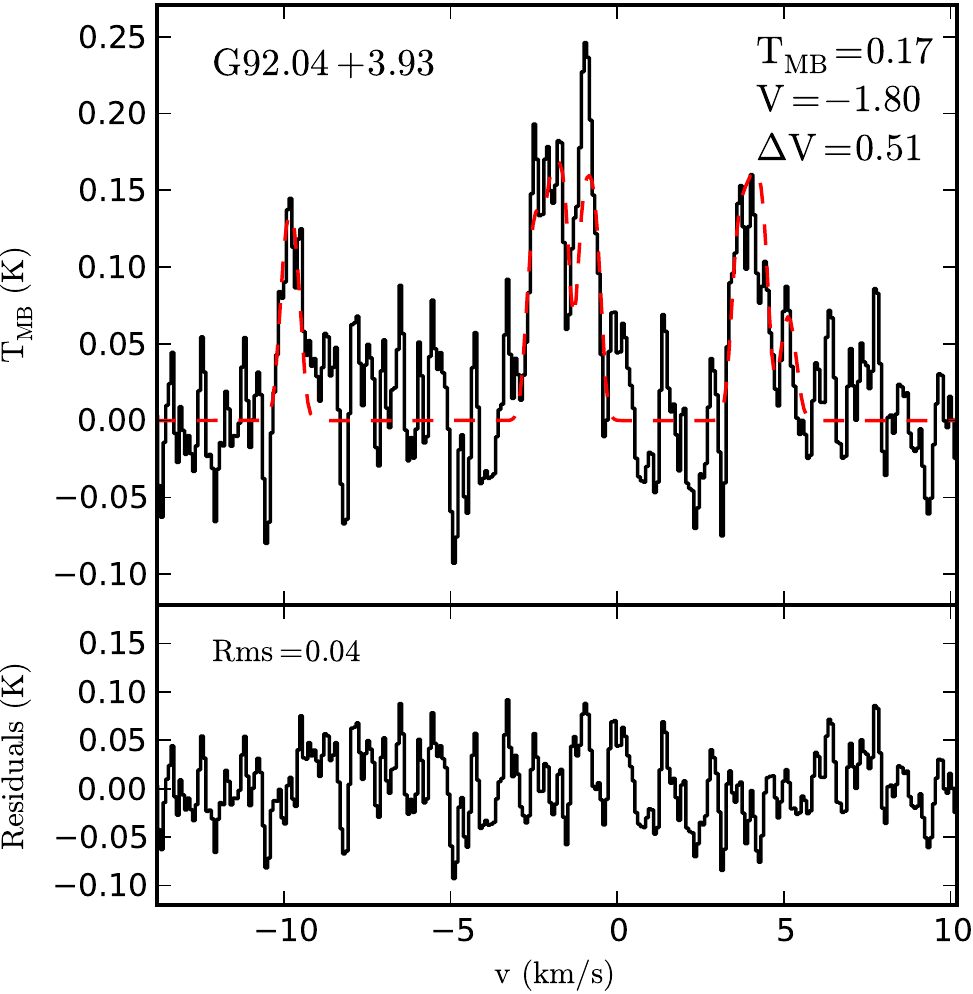}
        \includegraphics[scale = 0.69]{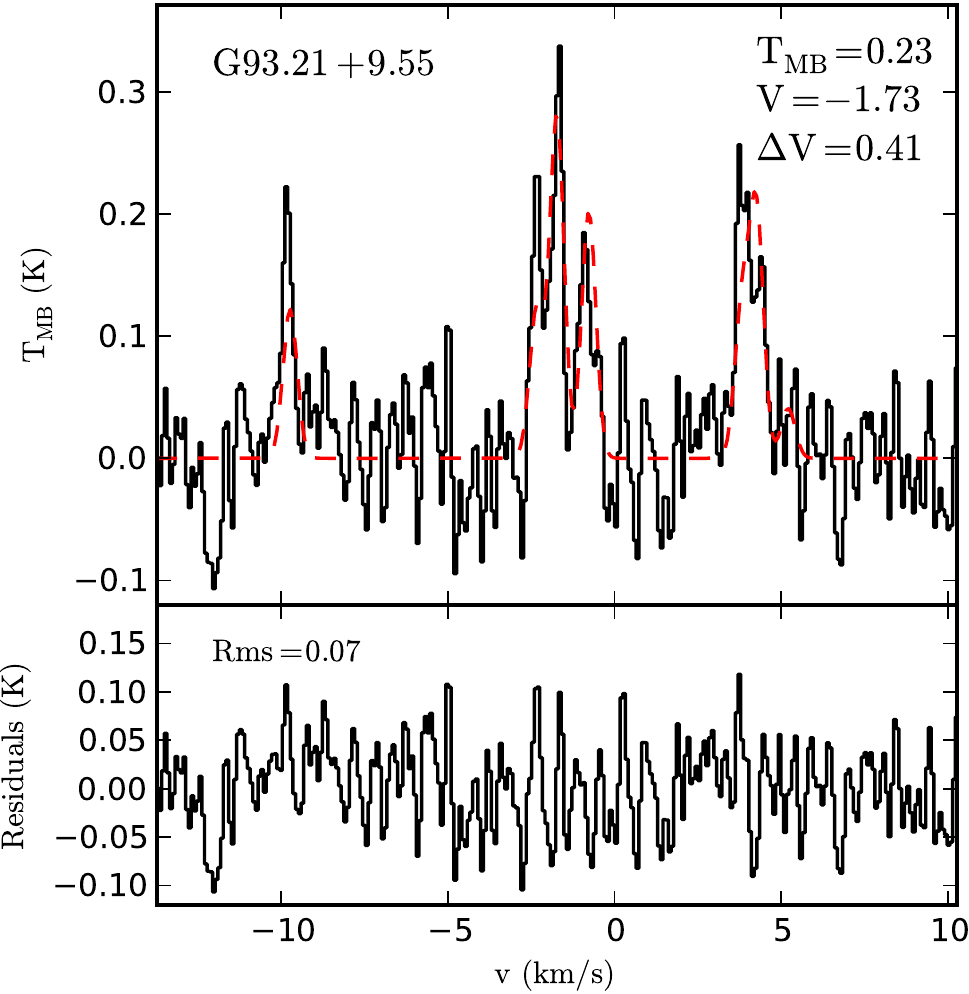}
        \includegraphics[scale = 0.7]{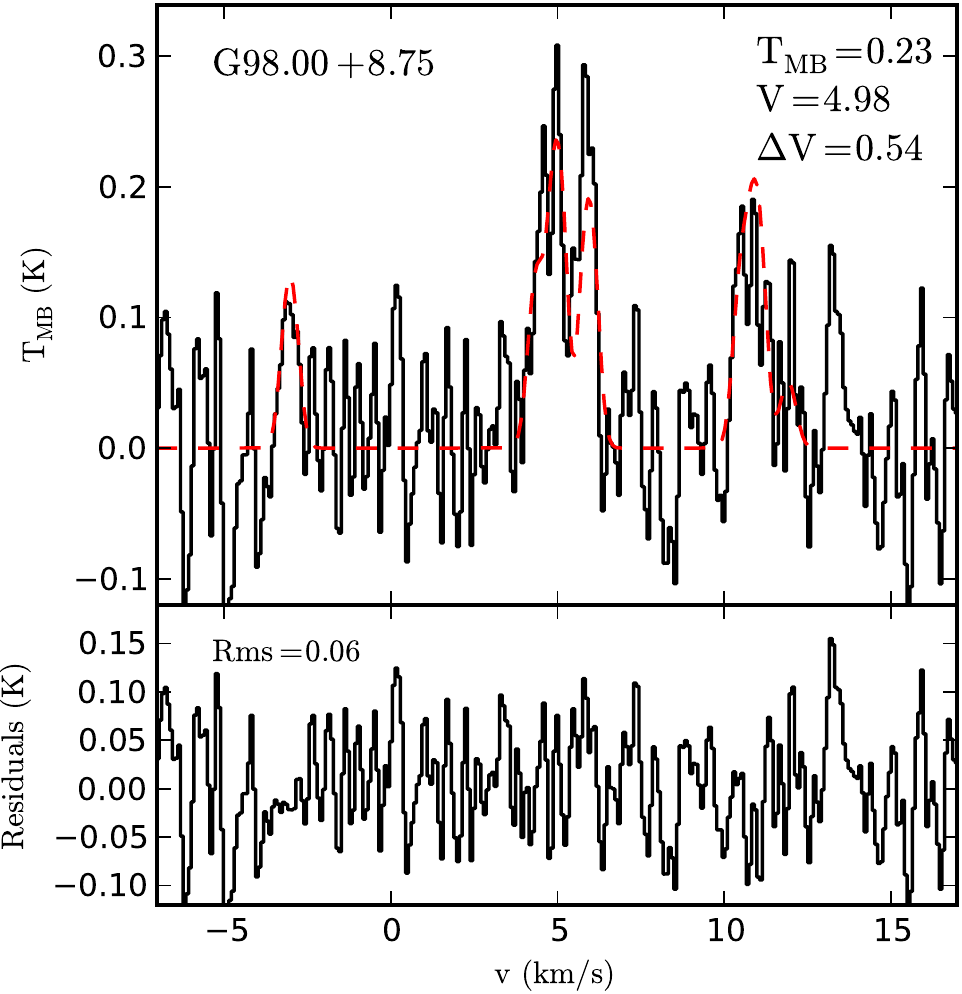}
        \includegraphics[scale = 0.7]{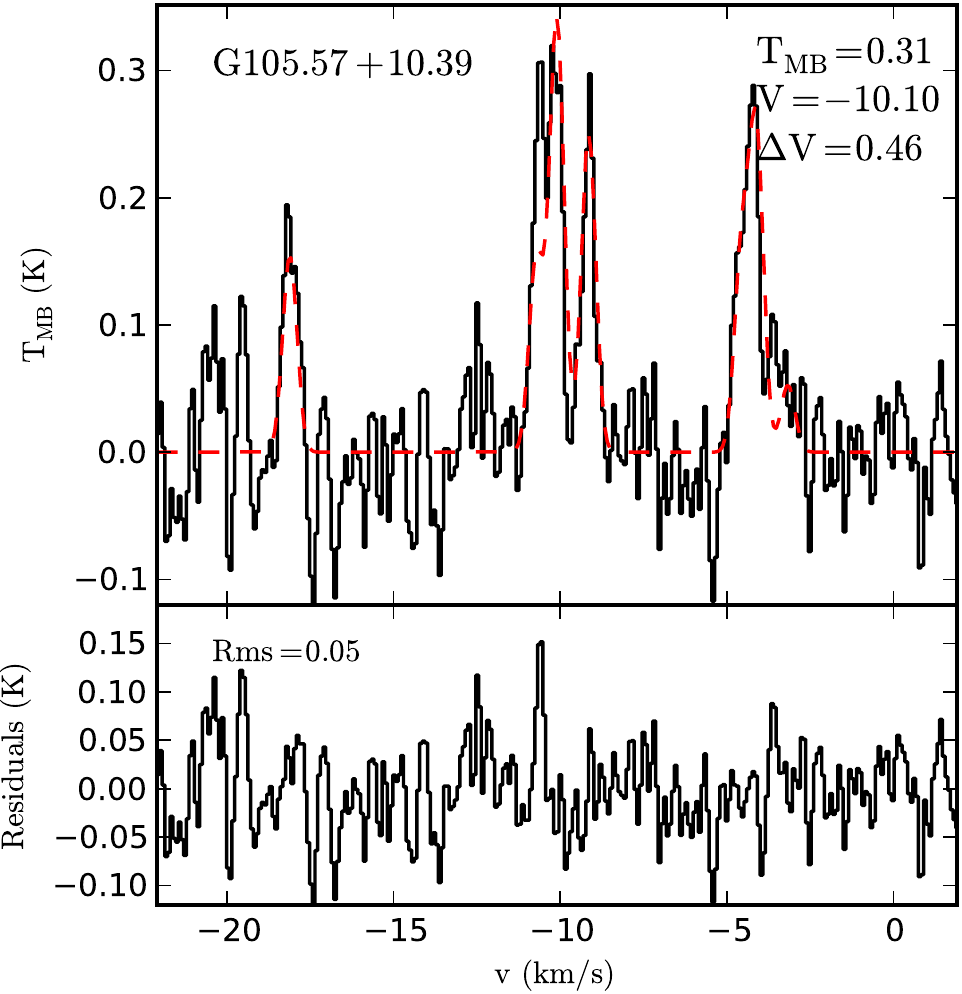}
        \includegraphics[scale = 0.7]{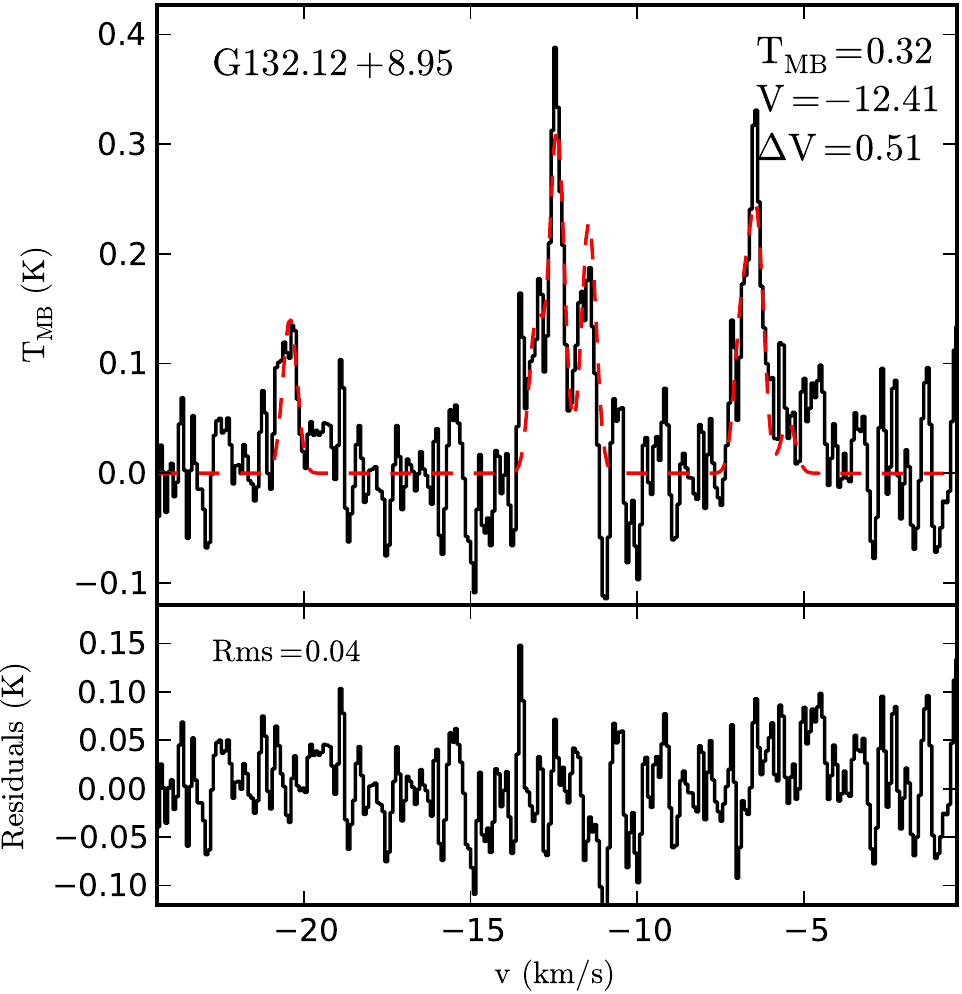}
    \caption[]{The ${\rm N_2H^{+}}$ spectra observed toward G92.04+3.93, G93.21+9.55, G98.00+8.75,  G105.57+10.39, and G132.12+8.95. The field G86.97-4.06 is shown in Fig. \ref{n2h+_g86}. The red dashed line is the fit to the hyperfine spectra and the lower subframe shows the fit residuals. The main beam temperature of the 23-12 component, the fitted radial velocity, the FWHM line width (km~s$^{-1}$), and the residual rms noise (as main beam temperature) are given in the frames.}
    \label{n2h+_app_pic}
\end{figure*}

The spectra with N$_2$H$^+$ detections, besides the field G86.97-4.06 in Fig. \ref{n2h+_g86}, are shown in Fig. \ref{n2h+_app_pic}.

\section{Radiative transfer models}
\label{rtm}

In Sect.~\ref{results} the dust emission was analyzed assuming a constant temperature along the line-of-sight (LOS), but this may underestimate the true column density of externally heated, optically thick clumps \citep{Malinen2011, Juvela2013}. Similarly, the $^{13}$CO(1--0) and C$^{18}$O(1--0) lines were analyzed using the LTE assumption, without the possibility of independent estimates of their relative abundances. To examine these questions, we carried out radiative transfer modeling where the density distribution was first derived from dust continuum observations and was then used as a basis for modeling the lines. The modeling was limited to the two positions in the field G131.65+9.75.

\subsection{Modeling of dust surface brightness}

We extracted $5\arcmin \times 5\arcmin$ continuum surface brightness maps centered on the selected positions and resampled the data on 3$\arcsec$ pixels. The maps are 100$\times$100 pixels in size and corresponding model clouds were constructed using a cartesian grid of $100 \times 100 \time 100$ cells. In the line-of-sight direction the density distribution was assumed to follow a Plummer-like \citep{Whitworth2001, Plummer1911} function $\rho(r) \sim \frac{1}{[|1+(r/R_{\rm flat})^2|]}$, with a central flat part with $R_{\rm flat}$ equal to 0.03~pc (not well resolved with the employed discretization). Corresponding to the apparent clump sizes in the plane of the sky, the FWHM of the density distribution was set to $\sim$17 pixels, which corresponds to $\sim 50\arcsec$ or a linear scale of 0.25~pc at the distance of 1070~pc. This may overestimate the size because of the effects of beam convolution and radial temperature gradients in the clumps. For this reason, and to check the general sensitivity to the LOS extent, we also calculated another set of models with FWHM at half of the value given above. Note that the clump size can also be significantly larger along LOS than in the plane of the sky. There is even some bias in this direction because the emission from elongated clumps and filaments becomes stronger when they are aligned along LOS.

We performed the calculations  iteratively, and on each iteration we solved   the dust temperature distributions and calculating model predictions of the surface brightness that we then convolved to the resolution of the observations. The calculations were carried out with a Monte Carlo radiative transfer program \citep{Juvela2005a}. The initial external radiation field corresponded to that of \citet{Mathis1983a}. The modeling was performed with two dust models. The first, in the following MWD, represents dust in normal diffuse interstellar medium \citep{Li2001}. The other one, in the following OH, was taken from \citet{Ossenkopf1994} and corresponds to dust that has coagulated and accreted thin ice mantles over a period of 10$^5$ years at a density 10$^{6}$~cm$^{-3}$. In our calculations, the dust opacities $\kappa$($250\muup$m) were 0.045~cm$^2$~g$^{-1}$ for MWD and 0.22~cm$^2$~g$^{-1}$ for OH, the value of Sect.~\ref{results} falling between the two. The value of $\kappa$ is crucial because it affects the column density, an important parameter of the subsequent line modeling.

In the plane of the sky, the column density corresponding to each map pixel was adjusted by comparing the observed 350~$\muup$m surface brightness with the model prediction. The external radiation field was adjusted so that finally  the 160~$\muup$m and 500~$\muup$m predictions also agreed with the observations to within 10~\%. The observations could be fitted well with both dust models, with an external radiation field 70~-~80~\% of the \citet{Mathis1983a} values. Because we used background-subtracted surface brightness data, the model represents only the inner parts of the cloud without the diffuse envelope. Therefore, the radiation field in the models is somewhat weaker that the full radiation field outside the cloud. The column densities are listed in Table~\ref{table:modeling}.

The values obtained with the OH dust model are close to the column densities estimated in Sect.~\ref{results}. With MWD, the column densities are higher and in the northern point by more than the ratio of dust opacities, $\sim$5. For a given dust opacity and radiation field, there is a maximum surface brightness that can be produced with any column density. Thus, for a too small value of $\kappa$, the column density of the model might be grossly overestimated. When estimated with the NICER method (using 2MASS stars, a spatial resolution of two arcminutes, and a value of $R_{\rm V}$=3.1), the visual extinction of northern clump is less than 3~mag. Taking  the difference in the resolution into account, this is still consistent with the Sect.~\ref{results} estimates and the result from the models with OH dust. However, the $A_{\rm V}$ measurement appears to rule out the values obtained with MWD dust (assuming the $A_{\rm V}$ is not severely underestimated either because of very clumpy column density structure or the presence of foreground stars) and give some support to the idea of dust opacity higher than that of diffuse medium.

\subsection{Modeling of the $^{13}$CO(1--0) and C$^{18}$O(1--0) lines}

We modeled the $^{13}$CO(1--0) and C$^{18}$O(1--0) lines  separately,  taking the density distribution directly from the continuum models. We only modeled the spectra toward the center of the clumps. The fractional abundance and kinetic temperature were assumed to be constant but in the non-LTE models the excitation temperature does vary and gives more weight to the dense regions. The velocity field was initialized by giving each cell a random velocity vector with $\sigma_{\rm 3D}=1.0$~km~s$^{-1}$ and assuming a turbulent line width with Doppler velocity $\Delta v_{\rm D}=\sqrt{2} \sigma_{\rm 1D}=1.0$~km~s$^{-1}$. Thus the initial "microturbulence" within the cells is slightly larger than the "macroturbulence" between the cells. In the actual calculations, to match the observed line widths, both velocity components are scaled by the same number, typically by $\sim$0.3~-~0.4, and the thermal line broadening is added to the turbulent line widths. The predictions of the models are optimized regarding the line width (the scaling mentioned above) and the line intensity that is adjusted by changing the molecular abundance. There are four cases for each position corresponding to two different assumptions of the LOS cloud extent  and the two different dust models. Furthermore, some models  were also calculated with a kinetic temperature of $T_{\rm kin}=12.0$~K, instead of the default assumption of $T_{\rm kin}=10.0$~K. Again, the true kinetic temperature is unknown, but the comparison between $T_{\rm kin}=10.0$~K and $T_{\rm kin}=12.0$~K gives some idea of the associated uncertainties.

We performed the line calculations  with the Monte Carlo program described in \citet{Juvela1997a}. The obtained abundances are listed in Table~\ref{table:modeling}. Concentrating on the models based on continuum modeling with OH dust, the $^{13}$CO values are on the order of the canonical value of $10^{-6}$ and the estimates are similar for both assumed values of $T_{\rm kin}$. However, because the abundances directly depend on the assumed column density, and thus the values of dust $\kappa$, the absolute values are very uncertain. The relative abundance between $^{13}$CO and C$^{18}$O should be more reliable, as suggested by the similarity between the OH and MWD cases and the two cases of LOS density distribution. The abundance ratio is $\sim$10 and thus larger than the value of 5.5 that was assumed in Sect.~\ref{methods}.

\begin{table*}
\centering
\caption[]{
Result from radiative transfer modeling. The columns are (1) observed position, (2) column density estimated in Sect.~\ref{results}, (3) width of the LOS density profile, (4) dust model, (5) column density of the continuum model, (6) assumed kinetic temperature, (7) $^{13}$CO abundance of the model, and (8) relative abundance [$^{13}$CO]/[C$^{18}$O].
}
    \begin{tabular}{llllllll}
    \hline
    Position  & N(H$_2$)  & LOS width & Dust  & N(H$_2$)    &  $T_{\rm kin}$  &  [$^{13}$CO] &  [$^{13}$CO]/[C$^{18}$O]  \\ 
              & (10$^{21}$cm$^{-2}$) &  &   &  (10$^{21}$cm$^{-2}$)   &  (K) &   (10$^{-6}$) &   \\ 
    \hline     
    G131.65+9.75 N &  9.3 &   Wide &  OH & 10.9 & 10.0 &   0.36 &   11.4 \\  
    G131.65+9.75 N &  9.3 & Narrow &  OH & 12.3 & 10.0 &   0.32 &   11.3 \\  
    G131.65+9.75 N &  9.3 &   Wide & MWD & 85.8 & 10.0 &   0.04 &   10.4 \\  
    G131.65+9.75 N &  9.3 & Narrow & MWD & 93.3 & 10.0 &   0.04 &   10.5 \\
                   &      &        &     &      &      &        &        \\
    G131.65+9.75 S &  3.7 &   Wide &  OH &  3.4 & 10.0 &   1.50 &   16.2 \\
    G131.65+9.75 S &  3.7 & Narrow &  OH &  3.7 & 10.0 &   1.22 &   14.8 \\ 
    G131.65+9.75 S &  3.7 &   Wide & MWD & 20.3 & 10.0 &   0.12 &   10.7 \\  
    G131.65+9.75 S &  3.7 & Narrow & MWD & 22.2 & 10.0 &   0.11 &   10.8 \\
                   &      &        &     &      &      &        &        \\   
    G131.65+9.75 N &  9.3 &   Wide &  OH & 10.9 & 12.0 &   0.32 &   10.7 \\  
    G131.65+9.75 N &  9.3 & Narrow &  OH & 12.3 & 12.0 &   0.29 &   10.8 \\  
    G131.65+9.75 S &  3.7 &   Wide &  OH &  3.4 & 12.0 &   1.11 &   13.3 \\ 
    G131.65+9.75 S &  3.7 & Narrow &  OH &  3.7 & 12.0 &   0.94 &   12.7 \\  
\hline             
\hline
\end{tabular}
\label{table:modeling}
\end{table*}

\begin{figure}
\centering
\includegraphics[width=8cm]{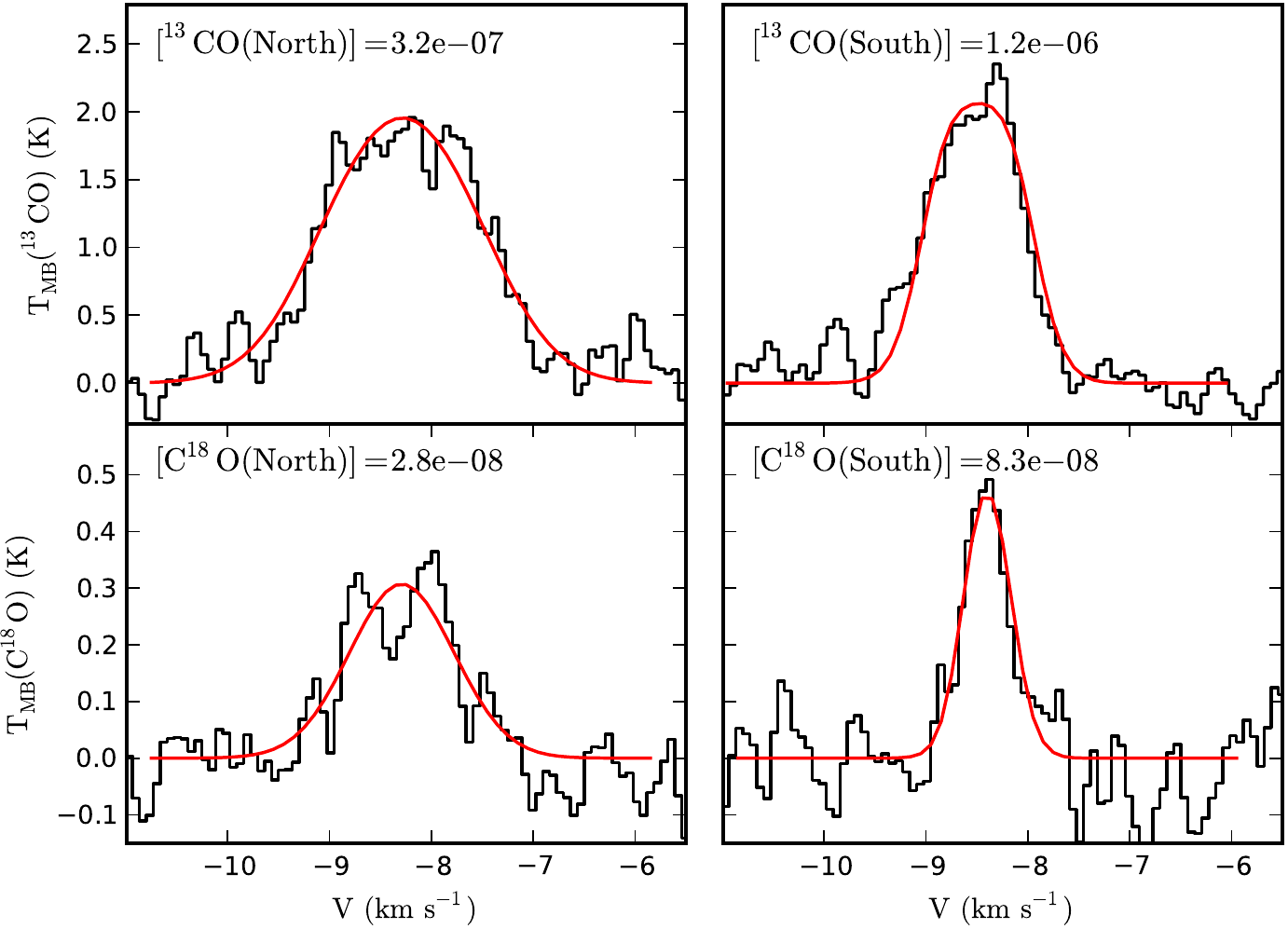}
\caption{
Examples of the modeled $^{13}$CO (upper frames) and C$^{18}$O spectra (lower frames) for the northern (left frames) and southern (right frames) positions in the field G131.65+9.75. The histograms show the observed spectra and the continuous red lines the model predictions.
The model densities were derived from continuum modeling with OH dust, assuming a narrow LOS density profile.
}
\label{fig:sample_spectra}%
\end{figure}

\bibliographystyle{aa}
\bibliography{AA201423428}

\end{document}